\documentclass[sigconf]{acmart}

\copyrightyear{2022} 
\acmYear{2022} 
\setcopyright{rightsretained} 
\acmConference[CHI '22]{CHI Conference on Human Factors in Computing Systems}{April 29-May 5, 2022}{New Orleans, LA, USA}
\acmBooktitle{CHI Conference on Human Factors in Computing Systems (CHI '22), April 29-May 5, 2022, New Orleans, LA, USA}
\acmDOI{10.1145/3491102.3517517}
\acmISBN{978-1-4503-9157-3/22/04}

\AtBeginDocument{%
  \providecommand\BibTeX{{%
    \normalfont B\kern-0.5em{\scshape i\kern-0.25em b}\kern-0.8em\TeX}}}

\usepackage{graphicx}
\usepackage{xcolor}
\usepackage{verbatim}
\usepackage{pdfpages}
\usepackage{dirtytalk}

\begin{document}

\title[``You have to prove the threat is real"]{``You have to prove the threat is real":  Understanding the needs of Female Journalists and Activists to Document and Report Online Harassment}

\author{Nitesh Goyal}
\authornote{Work done while at Jigsaw, Google}
\email{teshg@google.com}
\orcid{0000-0002-4666-1926}
\affiliation{
  \institution{Google Research, Google}
  \city{New York}
  \country{USA}
}

\author{Leslie Park} 
\authornote{Work done at Google via Genesis10}
\email{ljpark.us@gmail.com}
\orcid{0000-0001-9305-5569}
\affiliation{%
  \institution{Genesis10}
  \city{New York}
  \country{USA}
}

\author{Lucy Vasserman}
\email{lucyvasserman@google.com}
\orcid{0000-0002-6938-0713}
\affiliation{%
  \institution{Jigsaw, Google}
  \city{New York}
  \country{USA}
}

\renewcommand{\shortauthors}{Goyal, et al.}

\begin{abstract}
Online harassment is a major societal challenge that impacts multiple communities. Some members of community, like female journalists and activists, bear significantly higher impacts since their profession requires easy accessibility, transparency about their identity, and involves highlighting stories of injustice. Through a multi-phased qualitative research study involving a focus group and interviews with 27 female journalists and activists, we mapped the journey of a target who goes through harassment. We applied the existing PMCR framework, as a way to focus on needs for Prevention, Monitoring, Crisis and Recovery. We focused on Crisis and Recovery, and designed a tool to satisfy a target's needs related to documenting evidence of harassment during the crisis and creating reports that could be shared with support networks for recovery. Finally, we discuss users’ feedback to this tool, highlighting needs for targets as they face the burden and offer recommendations to future designers and scholars on how to develop tools that can help targets manage their harassment.
\end{abstract}



\maketitle

\section{Introduction}

In 2017, two journalists - one in India, and one in Malta - were found independently murdered after having received repeated online harassment and threats by criminals who were reported by these journalists \cite{posetti2017conversation}. Online harassment is a real issue that has real world offline implications. Yet, online harassment remains a notoriously widespread threat, rapidly expanding and evolving globally and across platforms. As Internet adoption and online communities continue to grow, there is an imminent interest in understanding how online harassment develops and proliferates. That has led to us discovering that it can have detrimental impacts to an individuals’ mental health \cite{stevens2021cyber}, physical safety \cite{ojanen2015connections, williams2020hate}, gender equality in journalism \cite{posetti2020online, posetti2018violence, UNESCO2021}, and free press \cite{carlson2020online}. Further, targets of online harassment often face challenges in responding to attacks due to social shame, victim blaming, and silencing tactics as well as reaching the right support resources \cite{lumsden2017media, veletsianos2018women, jane2017gendered, sugiura2020victim, jane2020online, younas2020patriarchy}.

Since online harassment has become a pervasive part of digital life, a growing body of research continues to understand how different demographic factors like gender, age, and sexual orientation  etc. might lead to different forms and levels of harassment \cite{thomas2021sok}. Research has shown that women, people of color, LGBTQ, and younger individuals are more likely to be targets of online harassment \cite{scheuerman2018safe, blackwell2017classification, duggan2017online, vincent2020}. The literature highlights that women often report experiencing frequent, severe, and sexualized forms of online harassment \cite{seralathan2016making, marshak2017online, chen2018nastywomen, jane2017gendered, sambasivan2019they}. Studies have raised concerns around women’s ability to participate in online public spaces, as women often reported feeling desensitized to online harassment and employing strategies to manage access to themselves (e.g. self-censorship or avoiding interactions online) to combat online harassment \cite{chadha2020women, vitak2017identifying, fox2017women, backe2018networked, women2020online}. The public, and sometimes ``controversial,'' nature of journalists' profession attribute to high predisposition to facing online harassment \cite{holton2021not, arora2020novel, belair2019audience, chen2020you, lewis2020online, lenhart2016online}.

These trends are further compounded for female journalists \cite{UNESCO2021, arora2020novel, holton2021not, lewis2020online, chen2018nastywomen, chen2020you, posetti2018violence, antunovic2019we}. In late 2020, UNESCO and the International Center for Journalists (ICJ) conducted a global survey about online violence against women journalist and found that 73\% of women had experienced online harassment; 25 \% and 18 \% experienced receiving threats of physical/sexual violence; 20 \%  being attacked or abused offline; and 26 \% reported negative impacts to mental health \cite{posetti2020online}. Further, 38\% made themselves less visible online, 30\% self-censored on social media, 20\% withdrew from all online interaction, and 18\% specifically avoided audience engagement. Furthermore, 11\% reported missing work, 4\% reported quitting their jobs, and 2\% reported abandoning journalism altogether. In 2016, 10\% of women journalists said that they had considered leaving the profession out of fear, that agrees with data from other studies \cite{arora2020novel, carlson2020online}. Defeated and desensitized to online harassment, women journalists shared that their experiences with severe and frequent forms of harassment are ``a thing we just have to accept.'' \cite{holton2021not}.

While a recent scoping review about online abuse against female journalists revealed an emerging body of literature that demonstrated the research community’s growing recognition and interest in the issue, there exists a research gap in designing and building tools that can empower targets of online harassment \cite{simoesonline}. Similarly, while the community has paved the way to help targets of harassment in many meaningful ways, from building theoretical frameworks to help targets of online harassment during moderation processes \cite{sarita2021} to building tools that can help them take screenshots of one-off harassment \cite{sultana2021unmochon} with help from admins - a gap still exists to enable targets of harassment to manage harassment on social-media platforms at scale, in bulk and without needing significant support/access from admins or moderators. In particular, what if those targets are journalists or activists, like the ones in Malta or India, who can not turn off their social media accounts to remain accessible to general public. To address this gap, this work answers the following research questions:

\begin{itemize}
\item {How does the trajectory of an online harassment look like for a journalist/activist? What happens before, during and after an attack?
}
\item {What are the challenges faced by journalists/activists during this trajectory and what needs related to documentation/reporting remain unmet with the current situation?}
\item {How can we design tools to meet these needs and challenges?
}
\end{itemize}

These research questions are addressed in this paper as we walk the readers through our journey of identifying needs of female journalists' and activists' using focus group and interviews, understanding stages of harassment, and giving agency to these targets to manage their online harassment using a newly designed tool. Hence the primary contributions of this work includes the following:

\begin{itemize}
\item Understand what happens before, during, and after an attack on target of harassment
\item Highlight needs during the stages of harassment by applying the PMCR (Prevention, Monitoring, Crisis, and Recovery) theoretical framework \cite{jigsawmedium} 
\item Present a prototype addressing a key crisis and recovery need: documenting evidence and reporting harassment
\item Recommend design directions for community to explore the space for managing harassment
\end{itemize}

\section{Background}
\subsection{Online harassment}
Online harassment is a broad and expansive field, as online harassment takes many forms and definitions \cite{ashktorab2016designing, seralathan2016making, backe2018networked, gutierrez2019classification, citron2014addressing, blackwell2017classification}. There is no standard agreed-upon definition of what online harassment entails \cite{jhaver2020identifying, lenhart2016online}. For the purposes of this paper, we refer to online harassment as using language that is targeted at an individual, by an individual/group of perpetrators, leading to target’s lesser participation in online conversations.

The connection between online and offline harassment has been studied by examining the connection between cyberbullying and face-to-face bullying \cite{ojanen2015connections}, the connection between instances of online speech and offline violence and extremism \cite{williams2020hate}, and the connection between online harassment and physical violence \cite{sambasivan2019they}. \citet{williams2020hate}'s analysis of police crime, census, and Twitter data found a positive correlation between  consumption of racially motivated hate speech and an increased likelihood to engage in online harassment, as further validated by  \citet{sambasivan2019they}'s interviews with 199 women and six NGO staff in south Asia.

Since transition from online to offline harassment has increasingly involved social media, research has shown an invested interest in understanding  observed activities and affordances on social media that perpetuate  progression of harassment from the digital world to the physical world \cite{steinmetz2020, chandaluri2019cross, redmiles2019just, vashistha2019threats}. Additionally, since experiencing harassment online or offline can carry social stigma and shame, targets of harassment prefer to remain anonymous while seeking social support and sharing their accounts \cite{andalibi2016understanding, nova2019online, moitra2021parsing, sambasivan2019they}. So, privacy and data sharing preferences are important considerations when designing for targets of harassment \cite{yoo2021anshimi}.

\subsection{Journalists' Experiences, Needs, and Challenges}
Journalists are prone to online harassment as having and maintaining a social media presence has become a professional expectation in the news media \cite{holton2021not, arora2020novel, belair2019audience, lewis2020online, chen2018nastywomen, lenhart2016online, posetti2018violence}. In a recent survey conducted by the Committee to Protect Journalists (CPJ), 90\% of American journalists described online harassment as the biggest threat facing journalists today, with women and minority journalists being disproportionately targeted online \cite{arora2020novel}. A recent study with more than 30 journalists also shared sentiments that many online harassment incidents are unreported and they are expected to independently manage incidents and this was an expected challenge in their profession \cite{holton2021not}. Further, authors found that women journalists shared more experiences and concerns about their experiences with chronic and escalatory harassment. 

This is a widespread issue. In a study involving 75 interviews with female journalists with experience working in Germany, Taiwan, the UK, and the US, participants reported frequently facing comments that criticized, attacked, marginalized, stereotyped, or threatened them regarding their gender and sexuality \cite{chen2018nastywomen}. Online harassment targeting women has widespread repercussions impacting not only their safety and well-being but also their freedom of expression and journalistic productivity \cite{posetti2018violence, posetti2020online, carlson2020online, UNESCO2021}. Journalists are commonly targets for online harassment, but many of their needs are under reported. An interview study with 17 journalists from New York City found that using automated and manual tweet deletion tools was common practice, and one of the main reasons journalists used these tools was to remove content containing online harassment from their feeds - however, such tools are limited in scale, volume and efficacy \cite{ringel2020proactive}.

\subsection{Need for evidence to manage online harassment}

Despite the commonalities between online and offline harassment, the nuanced differences have posed challenges to addressing online harassment incidents through legal systems and law enforcement. Technological advances have introduced changes, outpacing and outgrowing laws. Crimes committed in cyberspace have extensive social, political and economic implications and may ultimately challenge traditional criminal laws \cite{hamin2018cloaked}. Most models of criminal justice seek to identify and punish offenders. However, these models break down in online environments, where offenders can hide behind anonymity and lagging legal systems \cite{blackwell2018online}. 

According to Heart Mob's resources for helping targets understand their rights, ``The intent of cyberstalking is to cause harm or fear whereas the intent of cyber harassment is to annoy or torment" \cite{heartmob}. In either case,  building a case for legal recourse against online harassment requires evidence and data. While online harassment can create a trail of data points, strong evidence is needed that requires collecting all these data points about actions and actors involved, validating the veracity of these incidences, and documenting and creating these reports that can then be shared with authorities to pursue legal recourse or actions on behalf of the targets. As of now, this task is left upon the targets of harassment, who with limited technical knowledge, are not always fully equipped to collect all the evidence. Even when such evidence is collected, mostly in the form of screenshots, it can be easily tempered with \cite{zheng2019survey}.

This paper describes designing a tool for targets to create these reports such that they can be shared by them with their trusted network, with direct meta-data from social media platforms - making it harder to temper with, easier to manage than taking screenshots of a flowing screen where data can scroll up too fast to capture, and at larger scale and volume. However, our work is not the first one - many other solutions have also been designed.

\subsection{Related Works - Solutions for managing online harassment}
\subsubsection{Content Moderation}
Online harassment has the capacity to rapidly develop in scope and severity due to the internet’s interconnectedness. Many platforms have implemented content moderation practices to manage the scale of online harassment, which tend to follow two types of reactive interventions: (1) detecting problematic content at scale or (2) relying on community members to report problematic content and follow community guidelines  \cite{seering2020reconsidering, gorwa2020algorithmic, gillespie2020content}. Common content moderation practices include training human moderators to review, flag, and remove user generated content that violates the organizations’ community guidelines, such as posts including hate speech or violent threats . Another approach is to rely on users in the community to report problematic content using report buttons and making community guidelines easily accessible on an interface. However, relying on humans to moderate content is practically infeasible due to the high volume of user-generated content online as shown by \citet{van2018automatic}. As the need for efficient content moderation workflows continues to grow, organizations have adopted automated solutions to facilitate content moderation workflows \cite{gillespie2020content, mahar2018squadbox, seering2020reconsidering, sarita2021}. For instance, in 2017, Google launched Perspective API, which is a free and openly available suite of machine learning models for detecting text with a high probability of containing toxicity, insults, profanity, identity attacks, threats, and sexually explicit language \cite{perspectiveapi}.

\subsubsection{Platform enforced strategies}
Since online harassment usually spreads throughout platforms, such as social media sites and forums, many platforms have not only adopted automated detection mechanisms aimed at content removal but also enforced policies and procedures for blocking or suspending accounts associated with abusive behaviors \cite{mathew2019thou}. Many platforms, like Twitter, also enable users to block comments or limit access to their content from selected users \cite{ferrier2018trollbusters}. Further, Twitter has implemented block lists like Block Bot and Block Together to bulk block accounts from a community-curated or allegorically generated list of problematic accounts \cite{jhaver2018online}. As arbiters of the digital public sphere, platforms face the difficult task to balance protecting their users from online harassment and maintain users rights to free speech. 
Additionally, a critical challenge for integrating automated techniques is accounting for subjective human perspectives on what should be considered inappropriate and worthy of censorship \cite{scheuerman2018safe} as well as false positives in detecting abusive content or behaviors \cite{jhaver2018online}. 

\subsubsection{Tools tailored for targets}
Research community has also created tools that help targets by handling exposure to toxic content. 

In collaboration with Hollaback!, an advocacy organization tackling harassment issues, \citet{blackwell2017classification} collected design feedback throughout an iterative design process to build Heartmob, which is a private online community that provides online harassment targets with real-time support resources, such as witnesses and tools for intervening during an attack \cite{blackwell2017classification}. \citet{blackwell2017classification}’s work revealed that online harassment targets wanted support, such as assistance with documentation and reporting attacks to platforms - which is the focus of our work.

The Haystack Group at MIT Science and Artificial Intelligence Lab developed Squadbox, a tool designed to help targets manage online harassment received through email by allowing targets to organize ``squads'' of individuals from the targets’ support network to act as moderators  who can reject, organize, or collaboratively apply filters to emails that contain harassment \cite{mahar2018squadbox}. Squadbox heavily focuses on content moderation to address real-time emotional and safety needs rather than targets’ reporting and documentation needs.

BlockParty is another application using a pay-for-use model that helps targets experiencing online harassment on Twitter to set filters that collect filtered content into a ``Lockout'' folder for review at their pace \cite{BlockParty}. BlockParty was designed to focus on limiting and controlling targets’ exposure to online harassment content, which is a critical and timely need, especially when online harassment attacks tend to proliferate quickly; however, the current application is not designed to provide other options like documenting and creating evidence. 

Troll-Busters.com is a tool designed to help journalists, by providing targets with immediate resources for emotional support and crisis response during an attack \cite{ferrier2018trollbusters}. When a target experiences an attack and shares the URL where the harassment is happening, a crisis response team inundates the targets’ Twitter stream with positive and encouraging tweets. While Troll-Busters.com provides a real-time response including emotional and security support, the tool is not as focused on gathering what went wrong.

Most recently, Unmochon was designed as a tool to help female targets of online harassment in the global south to capture authenticated screenshots on Facebook 

\cite{sultana2021unmochon}. Unmochon advanced the documentation and reporting process for targets as well as uncovered and addressed the unique challenges around verifying the authenticity of screenshots. However, actionability of reports generated from Unmochon rely first, on significantly deep relationship between a target and Facebook Admin - not equally accessible to everyone; second on continued access to Facebook platform itself - an option not available when access is compromised. Third, this tool depends upon acceptance of screenshots, which as mentioned in the last section can be tempered with - rendering them not sufficient of an evidence always \cite{zheng2019survey}. Fourth, this assumes that data generated during attack will stay static and can be taken screenshot of. Targets of harassment facing large scale attacks are unable to take such screenshots swiftly enough.

\section{Method}
This work was informed by research done across two phases of qualitative research and one design phase:
\begin{itemize}
\item Exploratory Phase:  The first phase was the exploratory phase to best understand the landscape and flow of harassment. 
\item Design Phase : This phase involved converting findings and user needs from Phase 1 into a tool design that would next be validated.
\item Validation Phase: The third phase involved validating whether our design directions were indeed reflective of the user needs and user feedback on further design considerations.
\end{itemize}
The entire process has been reviewed by internal processes and guidelines at our organization. Next, we will give further details about how the data was collected, and analyzed.

\subsection{Participants}
In total 27 participants were recruited for the focus group (n=9) and follow-up interviews (n=18). Participants were primarily female journalists, human rights activists and members of Non Government Organizations/Not-For-Profit support networks. For specific details about each participant, please refer to the Table~\ref{tab:participants}. All participants were recruited using convenience and snowball sampling and we received verbal/written consent for participation. 

We conducted a focus group (n=9) with female journalists and human rights activists affiliated with USAID and American Bar Association (ABA). The focus group took place at the end of an event and lasted approximately 90 minutes. Participants were provided free lunch and refreshments, as an incentive to participate.

We also conducted interviews (n=18) with a mix of online harassment targets and their advocates to gather deeper insights about experiences and challenges in managing online harassment and to evaluate our prototype for an online harassment manager. Participants were provided 30\$ for their participation during the Exploratory Phase, and another 30\$ for their participation during the Validation Phase.

\begin{table*}
\centering
\caption{List of Participants \label{tab:participants}}
\centering
\resizebox{\textwidth}{!}{
\begin{tabular}{ | l | l | l | l | }
\hline
	Participation & ID & Profession & Context \\ \hline
	Focus Group & FG P1 & Print Journalist & Female (age 30-40), works in the US, English as first language \\ \hline
	 & FG P2 & Human Rights Activist & Female (age 30-40), works in the US, English as first language \\ \hline
	 & FG P3 & Former Journalist/Independent Consultant & Female (age 30-40), works in the US, English as first language \\ \hline
	 & FG P4 & Print Journalist & Female (age 40-50), works in Asia, English as first language \\ \hline
	 & FG P5 & Activist & Female (age 50+), works in the US, Chinese as first language \\ \hline
	 & FG P6 & NGO/NFP Co-ordinator & Female (age 50+), works in the US, English as first language \\ \hline
	 & FG P7 & Activist & Female (age 30-40), works in the US, English as first language \\ \hline
	 & FG P8 & NGO/NFP Partner & Female (age 50+), works in the US, English as first language \\ \hline
	 & FG P9 & NGO/NFP Member & Female (age 50+), works in the US, English as first language \\ \hline
	Interviews & 1 & NGO/NFP Start-up Founder, Reporter & Female (age 20-30), works in the UK, English as first language \\ \hline
	 & 2 & TV and Radio Journalist and Reporter & Female (age 30-40), works in Colombia, Spanish as first language \\ \hline
	 & 3 & Print Journalist, Academic Scholar, NGO/NFP Consultant & Female (age 30-40), works in France, French as first language \\ \hline
	 & 4 & Ex-Journalist, now Print Editor & Female (age 40-50), works in US, Arabic as first language \\ \hline
	 & 5 & NGO/NFP Founder and Manager & Female (age 50+), works in US, English as first language \\ \hline
	 & 6 & Print Journalist, Activist, & Female (age 30-40), works in Tunisia, Arabic as first language \\ \hline
	 & 7 & Former Journalist/Independent Consultant & Female (age 30-40), works in the US, English as first language \\ \hline
	 & 8 & Print Journalist, Documentary Producer, Writer & Female (age 30-40), works in the Turkey, Turkish as first language \\ \hline
	 & 9 & NGO/NFP CEO & Male (age 40-50), works in the UK, Italian as first language \\ \hline
	 & 10 & Academic Scholar, Engineer & Male (age 20 - 30), works in the US, English as first language \\ \hline
	 & 11 & NGO/NFP Manager & Female (age 50+), works in the US, English as first language \\ \hline
	 & 12 & NGO/NFP Data Director & Female (age 20-30), works in the US, English as first language \\ \hline
	 & 13 & Journalist and Technologist & Male (age 50+), works in South Africa, English as first language \\ \hline
	 & 14 & Ex-Print Journalist, now web-designer and developer & Female (age 50+), works in the US, Italian/Romanian as first language \\ \hline
	 & 15 & NGO/NFP Senior Researcher & Female (age 20-30), works in the US, English/Urdu as first language \\ \hline
	 & 16 & NGO/NFP Digital Manager & Male (age 30-40), works in the US, English as first language \\ \hline
	 & 17 & NGO/NFP Manager & Information withheld \\ \hline
	 & 18 & NGO/NFP Consultant & Female (age 30-40), works in the US, English as first language \\ \hline
\end{tabular}
}
\end{table*}

\subsection{Exploratory and Validation Phase Analysis}

\begin{figure*}[ht]
  \centering
  \includegraphics[width=\linewidth]{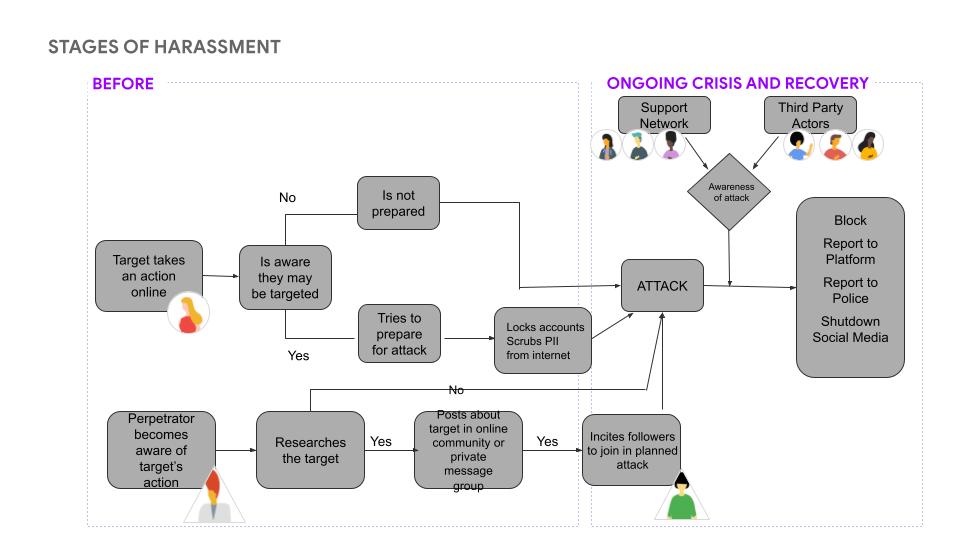}
  \caption{Stages of Online Harassment: Before, During, and After}
  \label{stages}
  \Description{Stages of Online Harassment: Before, During, and After an attack. A target may or may not be prepared for an attack while a perpetrator can attack by themselves or in collaboration with other followers.}
\end{figure*}

The goal of the focus group and interviews was to understand experiences with online harassment and identify emergent themes to inform a target's trajectory through the online harassment cycle. One of the authors created a research guide including semi structured questions. The questions included recent harassment experiences; flow of harassment - what happens before, during, and after; and reasons why targets feel that they were harassed. This work led to in Fig. ~\ref{stages}, described in further details over the next sections.

For both phases, video recordings were transcribed and one of the authors performed thematic analysis to produce an initial codebook. For Exploratory phase, codes were organized into three general categories: stages of harassment; needs of targets during each of these stages; and factors behind harassment. We applied an existing framework \cite{jigsawmedium} 
which we refer to as the PMCR framework throughout this paper. We chose this existing framework because we found that a chronological order of before, during and after harassment is not sufficient to view the needs distinctly. Instead needs cut across all stages of harassment: Prevention, Monitoring, Crisis, and Recovery (see also \cite{jigsawmedium}). We realized that themes cut across timeline of harassment - there is no single monolithic event called attack. Targets might be under one attack while recovering from a past attack or preparing for the next one. So, at a point of time - a person could be in one or more stages of harassment. 
Exploratory phase focus group and interviews yielded a codebook when the transcripts were analyzed using open coding. This led to a total of 20 themes and 25 codes, of which 6 codes reappeared in multiple themes. These findings discussed stages of harassment (before, during, after), needs of targets during these stages and the factors behind them. These findings were then classified into PMCR framework \cite{jigsawmedium}. 
Findings based on PMCR framework were then used to inform design of a new prototype.

Similarly for the Validation phase, the prototype was presented to 18 participants that have been marked as "interviews" in the previous section. The users were provided a overview of how the prototype was imagined to work; and how the data was captured, saved and could be operated upon. This took about 20 minutes. Next, participants were asked to think-aloud while trying to document/report a harassment episode that they had encountered in the last year. The task took about 25 - 30 minutes for them to recount their harassment episode and how they would use the prototype to find the harassment, document it and complete a report of that harassment episode. These sessions were audio-video recorded, transcribed, and analyzed using open coding. Overall we found 9 emerging themes and 21 codes.

Most salient themes are shared in the Exploratory and Validation Phase Findings section next.

\section{Exploratory Phase Findings}

\subsection{Trajectory of harassment and challenges faced by journalists/activists}
Analysis revealed that participants’ experiences with online harassment attacks included a distinct timeline of events typically occurring before, during, and after an attack incidence. Similarly, the online harassment perpetrators’ timeline involved a corresponding start, peak, and end of an attack. Sub-sections below detail participants’ experiences throughout the stages of an online harassment attack, as shown in Fig. ~\ref{stages}.

\subsection{Before an Attack}
\subsubsection{Publishing triggers attacks on journalists/activists}
\begin{quote}
``Usually what we see when a president or an elected official singles journalists, their loyal fans reply to this comment. Obviously, a comment of this elected official will be the most trending and toxic. Since the elected official facilitated this trolling in the first place, that's why when you see what's trending.'' - P16
\end{quote}

Perpetrators’ motivations for harassing journalists online can be construed from participants’ accounts of journalistic activities like publishing an article could trigger an attack. The scale of an attack can quickly escalate as a post gains momentum through high engagement and attention from dissenting individuals and groups. Perpetrators include trolls and individuals or groups with opposing political or social views from a journalist or publication. In some cases, influential figures like political leaders or government bodies can prompt an attack, since their online posts are often public and have a wide reach, as highlighted by P16.

\subsubsection{Alone, under-prepared, and bearing the burden to self-protect}

\begin{quote}
``Most people are totally lost, the average person has no idea [on what to do without an intermediary support person].'' 
- FG P3
\end{quote}

Despite their general awareness of harassment risks, many participants expressed concern and feeling under-prepared for a harassment crisis due to lacking knowledge of relevant resources and procedures for handling harassment attacks. The current state of affairs illustrates how the burden of preparing for and handling online harassment is placed primarily on at-risk targets and victims. This sentiment around under preparedness is echoed in the next section, which captures participants’ experiences during an attack.


\subsection{During an attack}

\subsubsection{Attacks on targets’ integrity via defamation}

\begin{quote}
``In relation to a Saudi Arabia story I wrote, some people sort of wanted to start this campaign against me to sort of denounce what I was writing.'' - P6
\end{quote}
\begin{quote}
``I was defamed by pro-government media, basically newspapers that belong to government officials or the ruling party. It was just like defamation after defamation. Then in addition to that online, there was this wave of harassment coming from the users. I couldn't obviously monitor everything.'' 
-  P8
\end{quote}

Perpetrators often try to discredit their targets. Tactics for discrediting included publicly denouncing journalists’ work, media-organized defamation, and creating false narratives about journalists, and  fabricating images of journalists. Making matters worse, affordances of online spaces facilitate the emergence and virality of online harassment, while reduce ability to manage the volume of the attacks

\subsubsection{Who and what can support a crisis response?}

\begin{quote}
 ``For the Jewish community, there are nationwide established security people in different cities. If I receive a threat, I contact [them]. They’re the ones who often interface with law enforcement, tech companies, or anyone else. They have a bigger voice [and] more bandwidth to deal with that more seriously.'' 
- FG P2
\end{quote}

In ideal scenarios, participants were able to work with their support network such as advocates, activists, community groups, news etc. Actors in participants’ support networks typically served as liaisons between the participants and specialists, such as platforms and law (e.g. law enforcement and lawyers), since activists and community groups have a greater voice and resources to advocate for participants. As is evident that managing an attack requires documenting and reporting information and resources that can be easily shared across the targets and their support networks - such a resource might must exist in a timely fashion for dealing with the attacks.

\subsubsection{Under-prepared for a crisis response}
\begin{quote}
 ``There needs to be more education for people to actually take precautions before the violence.'' - FG P6
\end{quote}

Several participants pointed out that navigating the right resources at the right time were challenges attributed to a general under preparedness to deal with online harassment. While some participants may be equipped with a resource list or support network as a precaution for harassment attacks, not all communities are well prepared to manage crisis and recovery needs, including documentation and reporting but also accessing the appropriate resources at critical times. 

\subsection{After an attack}
\subsubsection{Hard to collect evidence against organized attacks}
\begin{quote}
 ``There’s this one user on one platform and he’s on another now, and other platforms too...It takes weeks to get his social media accounts taken down. This person targets multiple people and has doxxed multiple people, but it’s really hard to document and report it as an organized effort.''
- FG P2
\end{quote}

After the peak of online harassment attacks, participants shifted their focus towards recovering, which involves documenting, reporting, and mitigating damages due to online harassment. Much of the recovery process continues from the attack peak or starts after the attack. At this stage, documenting online harassment incidents serves the important purpose of building evidence and credibility for reporting the attack. Since online harassment could entail complexities like sophisticated coordination or a rapid surge across various platforms, documenting incidents can be a difficult and taxing task.

\subsubsection{Documentation and reporting to publicly shame the harasser}
\begin{quote}
 ``Use media. Post about it, make it very public to help people understand what's happening and to discredit the attackers basically.'' - FG P3
\end{quote}

Due to challenges in the reporting process, one strategy is for participants to take matters into their own hands by trying to publicly discredit and/or combat harassment or documenting incidents on social media. However, this is a strategy that requires actively engaging with harassment, and perpetrators in a cycle where the targets have to use the platforms of attack as the venues to fight back against harassment as well. 

\subsubsection{Cycle of harassment fatigue}
\begin{quote}
 ``We’re seeing fatigue. If somebody wants to use online platforms to target me as an activist, they’re going to do it and it’s a little tiring to go out with this little bit of effort that I’m not sure is enough.'' - FG P8
\end{quote}

As a result, actionability of documentation remains a critical issue due to these shortfalls in current reporting systems that suffer from volumes of reports. Since reporting and recovery systems depend upon documenting harassment, and can require significant efforts , participants often grow de-sensitized to harassment, which creates a negative feedback loop into a system where online harassment perpetuates and reports can take time to be resolved . Multiple participants laughed in agreement to FG P8’s comment above. 
\begin{quote}
 ``At any stage of the process, we might totally disengage, get really burned out, and [feel] like ‘I don’t care anymore. Everybody knows everything about me. What do I have to hide?’ I don't know how many times I’ve heard that.''- FG P3
\end{quote}

These quotes illustrated a collective sentiment towards feeling fatigued and defeated in managing online harassment. This sentiment was also expressed in the before and during stages of the online harassment timeline. As is evident, the needs of targets of online harassment vary before, during and after an attack. In the next section we will discuss the PMCR framework \cite{jigsawmedium}
, a theoretical framework we applied to classify and categorize these needs.

\section{PMCR Needs Framework}

\begin{table*}
\centering
\caption{This table shows how data from our analysis during and after harassment map to the Prevention, Monitoring, Crisis and Recovery framework \cite{jigsawmedium} 
\label{tab:TimelineDuringAfter}}
\centering
\begin{tabular}{| p{1.9cm} | p{6cm} | p{6cm} | }
\hline
	Needs\textbackslash Timeline & During & After \\ \hline
	Prevention & ``It’s not just the one incident, it’s the consolation of different things happening on lots of different platforms.'' -FG P5 & ``It also depends on what else they have going on in their lives [...], but people don’t always have aspects of their life saved.'' -FG P9 \\ \hline
	Monitoring & ``When a journalist is trending, it’s probably because [they’re] being harassed or because something happened with a piece they published or [something] they did on TV or radio.'' - P2 & ``[There are unique challenges consolidating reports] to understand that’s a possible coordinated effort and not just individual users [attacking].'' -FG P2 \\ \hline
	Crisis & ``It's common practice [for perpetrators] to intimidate you by shutting windows or moving stuff around to freak you out. [After happening, I went to] Apple to secure my devices.'' -FG P9 & ``Use media. Post about it, make it very public to help people understand what's happening and to discredit the attackers basically.'' -FG P3 \\ \hline
	Recovery & ``If they truly feel like their life is being threatened, some [support organization] will walk and talk me through what my options are. For me, my go-to will always be to be in touch with someone who can start the process.'' -FG P7 & ``It took me a little over a year to go through the reports and what not. It’s a broken system where you don’t do anything about it, because it’s not worth my time.'' -FG P4 \\ \hline
\end{tabular}
\end{table*}

Our analysis revealed that targets of harassment do not think of harassment as a monolithic event - they think of harassment as an ongoing event. Needs of targets of harassment cut across multiple parts of harassment trajectory and we found that participants faced four types of needs regarding online harassment. These four needs aligned with the existing PMCR framework \cite{jigsawmedium} 
, which we summarize below and map to themes from our analysis and the online harassment timeline from the previous section.

\begin{itemize}
\item {Prevention: These needs include precautionary measures or knowledge for an online harassment attack -- what should I do to prepare or in case there is an attack?}
\item {Monitoring: These needs pertain to understanding the activities and landscape of online harassment -- what is happening? Am I being attacked?}
\item {Crisis: These needs involve the immediate response needed during an online harassment attack, such as addressing immediate safety/security compromises, finding the right resources, and figuring out how to handle the harassment attack -- How do I get this under control? What should I do now?}
\item {Recovery: These needs are about mitigating the online harassment impacts -- How can I recover from the attack?}
\end{itemize}

The PMCR framework aligned with our analysis on two parts of participants’ online harassment attack journeys: during and after. For more details please refer to the Appendix to see the mapping between the stages and PMCR. Despite participants’ general awareness of online harassment risks, participants reported feeling lost and underprepared during and after an attack. Based on participants’ accounts illustrated through these  key themes, it is evident that participants faced major problems with reaching resources, documenting, and reporting incidence during a crisis and recovering from them. 

As is also evident, participants brought up multiple other needs too like how does one prepare against an attack ? Or how does one monitor what is happening ? During the Design Phase and the rest of the paper, we focused solely on needs related to Crisis and recovery, and within that documentation and reporting, in particular, for multiple reasons. First, we wanted to narrow the scope of the work - so we had to make a choice. Second, we decided to choose Crisis and Recovery instead of Prevention and Monitoring because this is a societal problem about human behavior. Preventing harassment would require educating larger society. Monitoring harassment is a need that has significant security concerns for targets of harassment. Targets like journalists and activists are under constant surveillance. Designing another opportunity to monitor (even if they were in control of such tool) creates new security risks. Third, existing literature has shown that documentation and reporting can help, and is a gap that has clearly been identified \cite{blackwell2017classification, sultana2021unmochon}. So, we decided to focus energy on fixing the problem during and after it has occurred. Fourth, we have technology today that can detect toxic behavior after it has occurred and not before \cite{perspectiveapi}. So, we wanted to design a solution by leveraging such technologies that exist today and validate those designs.

Within Crisis and Recovery, users reported multiple needs like finding the right resources, and finding support. We decided to focus on Documentation and Reporting needs in particular because unless some tangible artifact exists as evidence - it is incredibly hard to provide support or fix the problem. This has been shown to be a gap in the existing design space as well \cite{sultana2021unmochon}. Hence, rest of this paper also focuses on one of the key themes: Documentation and reporting challenges, and associated user needs during and after an attack. Since documenting and reporting related challenges are faced during and after harassment (and not before harassment), we share some exemplary quotes mapped to PMCR needs during and after harassment in the Table \ref{tab:TimelineDuringAfter}

In the next section, we will unpack how these challenges and user needs informed the Design phase, followed by design of a tool that meets these needs. 

\section{Design Phase - understanding documentation and reporting challenges}
In this section, we will share specific data from Exploratory phase that has been used to inform our design process and the tool, presented in the next section.
\subsection{Taking screenshots is important to document and report harassment}
\begin{quote}
 `` Screenshots are always needed because you [need] evidence in case that account or that comment was suspended. Let's say when I'm writing about a campaign from a news website, we prefer to do a screenshot instead of linking directly to the news website, because that article could be taken down [and] we don't want to give those people page views. So I think screenshots are really useful. " - P6
\end{quote}

The interviews provided details that further validated the documentation and reporting challenges described in the focus groups. Documentation and reporting work in tandem to build credibility that the target experienced online harassment. Taking screenshots is a common form of gathering evidence about an attack. The evidence is used to support the target's reports to platforms or law enforcement. Screenshots often are also starting places for investigating who the perpetrators are, and what could be their potential motivations or sponsors?

\subsection{Content can be deleted, despite screenshots}
\begin{quote}
``A lot of the trolls would write nasty things and then they would delete it a day later or something. So it's difficult sometimes to track. Sometimes they delete, and I don't know who deleted it. If I see a really nasty comment, even if it's not addressed to me, I will report it myself. Then I'm like, ``Did I imagine that?'' So it's something that I think about, and it creates this unsafe environment emotionally for journalists.'' - P4 
\end{quote}

Online content is relatively easy to delete or be removed. Further complicating matters, perpetrators’ posts or comments containing online harassment can be deleted for many reasons: the perpetrator who wrote the post or comment deleted it, a bystander reported the perpetrators’ account or content, the platform automatically flagged the perpetrators’ account or content, or the target reported or removed the perpetrators’ content.  

The reason for content being deleted or removed online is often unclear, as content removals are not often well documented. For instance, reported content and content moderator’s decisions lack transparency. When content is no longer available online due to deletion, accessing it as a piece of evidence to build a case is challenging. Some journalists and activists even pointed to such deletions leading to self-doubts about validating the existence of attacks.

\subsection{Legal processes and data policies make it harder to gather deleted content}
\begin{quote}
``In the U.K., we've got GDPR, so it means that platforms can only hold that data for a certain amount of days. If you [want] to go to law enforcement, and they work quite slowly, you need proof from platforms to say that the comments you've pulled in your report are true and it's from the platform. Maybe there needs to be something around making it easy for a request of the data from the platform to be made.'' - P1 
\end{quote}

Online harassment is incredibly difficult to document and report, due to data retention policies and procedures to build a legal case. Targets are tasked with the burden of proof. Building a legal case could require proving negative psychological impact and the existence of a real threat to the target, as described in P1’s quote. An added nuance to proving online harassment is that the proof is expected to be documented. However accessing data can be difficult due to data retention policies like GDPR or as mentioned in the previous subsection, content can be easily deleted or removed. Further, how the proof is presented impacts the sensemaking processes of intelligence analysts that wish to provide support but are drowning under significant amount of data \cite{GIS2013, GIS2014} and their own cognitive biases \cite{GST2016,GCS2017}.


\subsection{Crisis response requires timely and human support from support networks}
\begin{quote}
``You're crying in front of your laptop, and you know what I want? I would like to get access to resources, external resources, third parties or [online] resources. Or advice like a security checkup to protect yourself better. Or [reassurance], "Okay, it's not a happy moment, but we're going to help you to end all this." - P3
\end{quote}




Participants also emphasized the need for emotional support while dealing with an online harassment attack. Overall, the interview sentiment reiterated the sentiment that participants’ experiences in dealing with online harassment are often overwhelming, confusing, and exhausting, especially when online harassment can trigger emotional and psychological distress as well as pose physical safety and security risks. In addition to practical support resources, participants expressed the need for human support and reassurance to guide them through the difficult experience.


\subsection{In-house/employer based support network is inequitable}

The prevalence of online harassment targeting journalists established a need for crisis response surveillance and resources for navigating threats and attacks. While large and established media organizations are most likely to have the resources and procedures in place to  support their staffed journalists throughout attacks, smaller media and non-profit humanitarian organizations do not face the same reality. 


After having identified the particular challenges related to documentation and reporting, our next step was to identify how do we design to satisfy some or all of these challenges. We distilled these findings into 5 Design Goals (referred to as DG 1 to 5 in the next section) by connecting each of these goals with one or more of the challenges highlighted above. We postulate that these are just some of the potential ways in which the challenges above could be met based on our brainstorming. We would encourage the readers to explore further options and evaluate them.

\section{Designing for documentation and reporting challenges}
As highlighted in the previous section, journalists/activists face particular challenges when documenting and reporting their harassment. While screenshots of harassment seems to be the most widely used way to document harassment, they pointed to screenshots not being useful anymore when the content gets deleted, which reportedly happens often. After deletion, there is no way to validate or verify the screenshots anymore either because the content can only be accessed by connecting directly to the back-end of the platform - an opportunity that does not exist for most of the journalists/activists. Based on the findings in the previous section, we identified the following design goals:
\begin{enumerate}
   \item DG1. Privacy Considerations: One of the privacy concerns that a design should manage is related to who can access what data ? Only the target should be able to access the tool, and see only their own data. Tool should not be used as a mechanism to attack others. This is table-stakes for any design as highlighted by previous literature we references about harassment carrying social stigma and shame, leading to targets of harassment preferring to remain anonymous while seeking social support and sharing their accounts \cite{andalibi2016understanding, nova2019online, moitra2021parsing, sambasivan2019they, yoo2021anshimi}.
  \item DG2. Data Validity: To overcome challenges related to validity, a tool should connect directly with the back-end of platforms for validating data even if the data is deleted after wards. (Based on Section 6.2 and 6.3)
  \item DG3. Data Analysis: Users should be able to choose/find particular information related to harassment across multiple dimensions. (Based on Section 6.3)
  \item DG4. Manage Volume: Users should be able to manage the results volume to easily create evidence in the middle of an ongoing volley of attack. (Based on Section  6.1 and 6.5)
  \item DG5. Share with Support Network: Users should be able to easily share gathered evidence with their support network for further help. (Based on Section 6.4)
\end{enumerate}

\begin{figure*}[!tbp]
  \centering
  \begin{minipage}[b]{0.48\textwidth}
    \includegraphics[width=\textwidth]{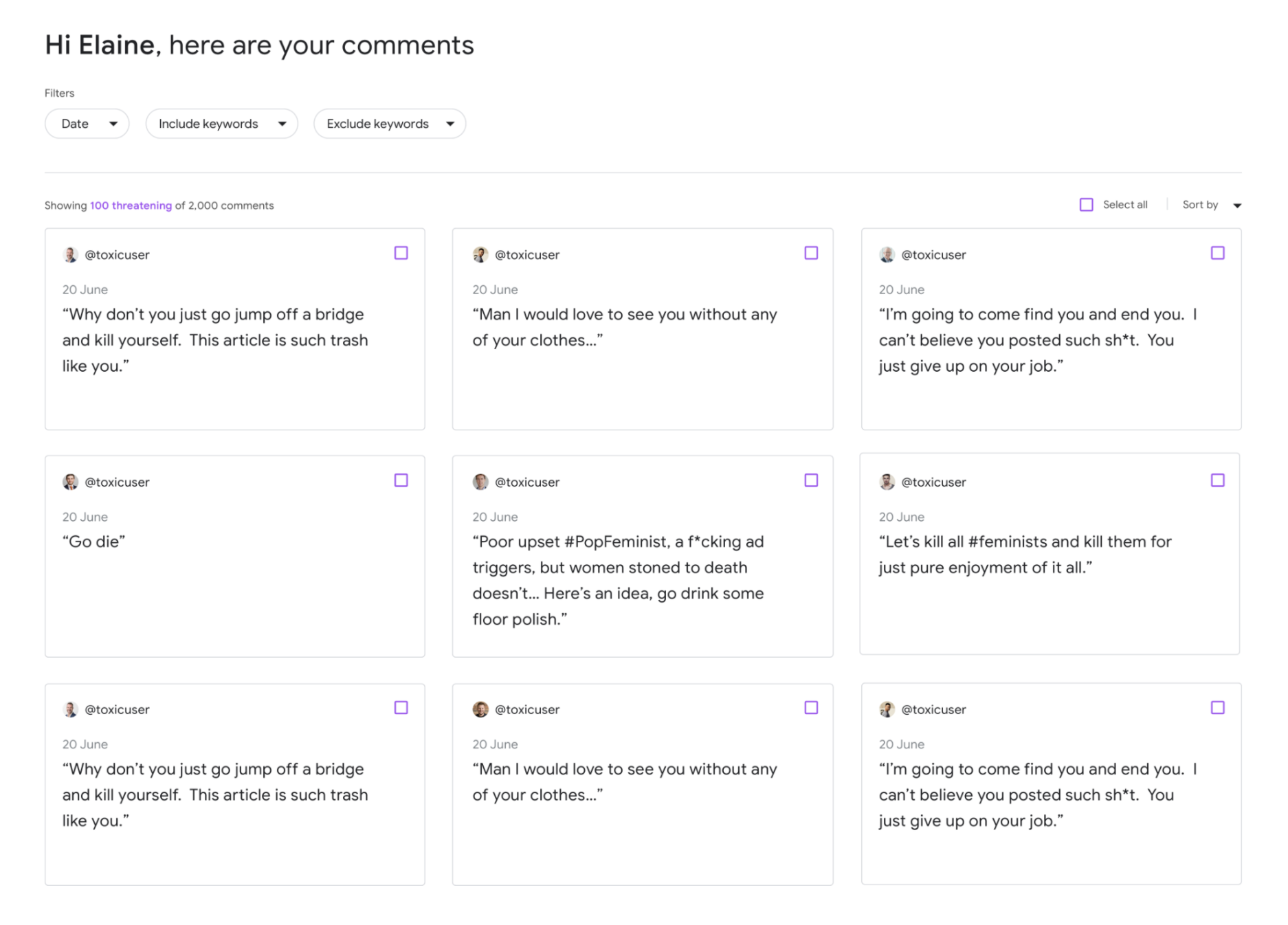}
    \caption{Homepage where a person can view all the comments that have been captured via connection with the back-end and the API}
          \Description{Homepage where a person can view all the comments that have been captured via connection with the back-end and the API}
    \label{fig:tool3}
     \hspace{-4mm}
   \end{minipage}
  \hfill
  \begin{minipage}[b]{0.48\textwidth}    
    \includegraphics[width=\textwidth]{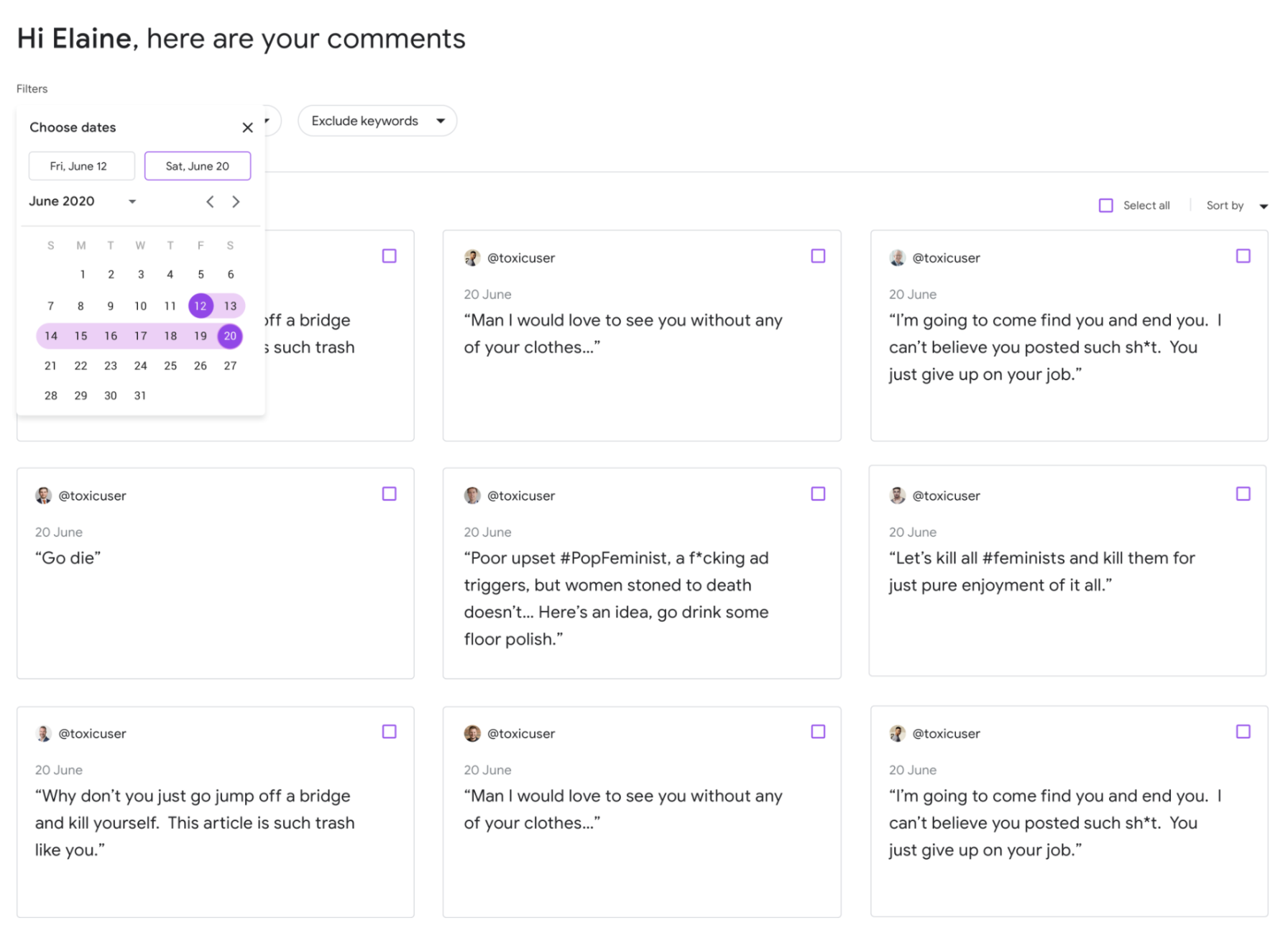}
\caption{Homepage where a person can choose to filter based on dates, including certain keywords/hashtags/usernames etc.}
    \label{fig:tool4}
        \hspace{-4mm}
  \end{minipage}
      \Description{Use filters like date etc. to identify the harassing text in one page}
\end{figure*}

\begin{figure*}[!tbp]
  \centering
  \begin{minipage}[b]{0.48\textwidth}
    \includegraphics[width=\textwidth]{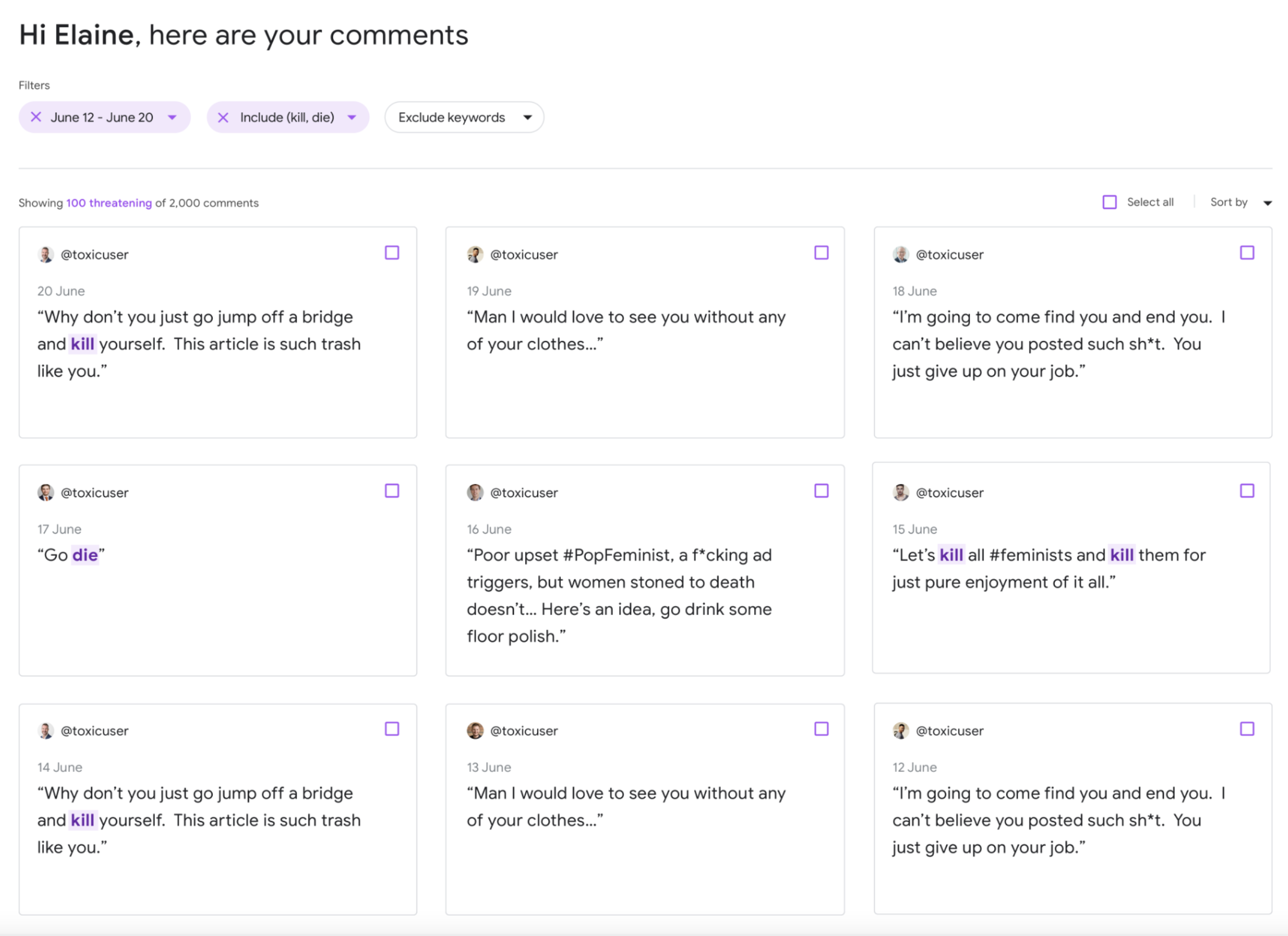}
    \caption{Aggregating Comments/Data from Platforms and using Filters to include/exclude data to manage volume}
        \Description{Aggregating Comments/Data from Platforms and using Filters to include/exclude data to manage volume}
    \label{fig:tool1}
    \hspace{-4mm}
  \end{minipage}
  \hfill
  \begin{minipage}[b]{0.48\textwidth}
    \includegraphics[width=\textwidth]{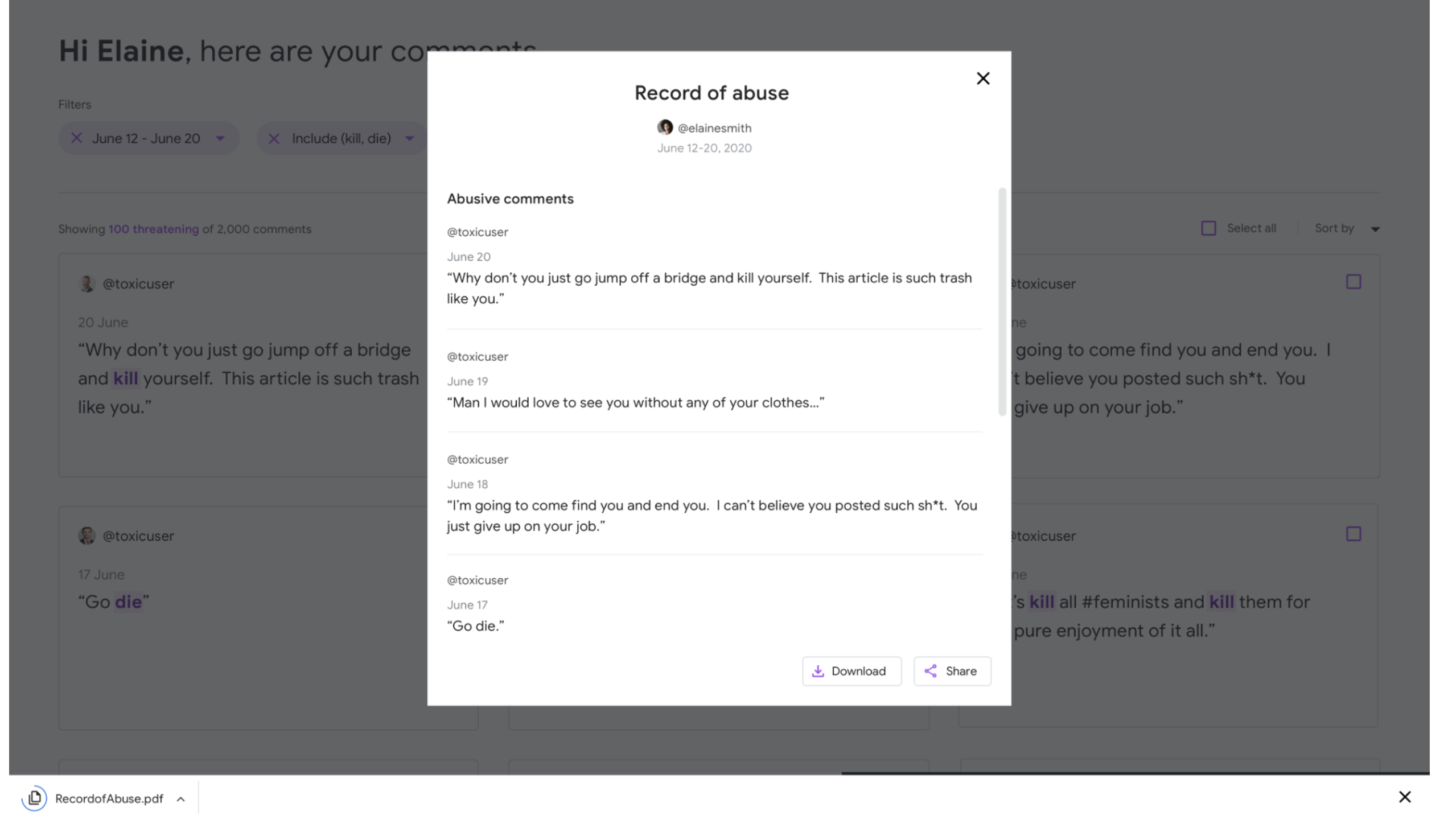}
    \caption{Adding aggregated data to a Report that can be downloaded or shared with support network}
    \label{fig:tool2}
    \hspace{-4mm}
  \end{minipage}
    \Description{View all the harassing text in one page and add them to a report}
\end{figure*}

\begin{figure*}[!tbp]
  \centering
  \begin{minipage}[b]{1\textwidth}
    \includegraphics[width=\textwidth]{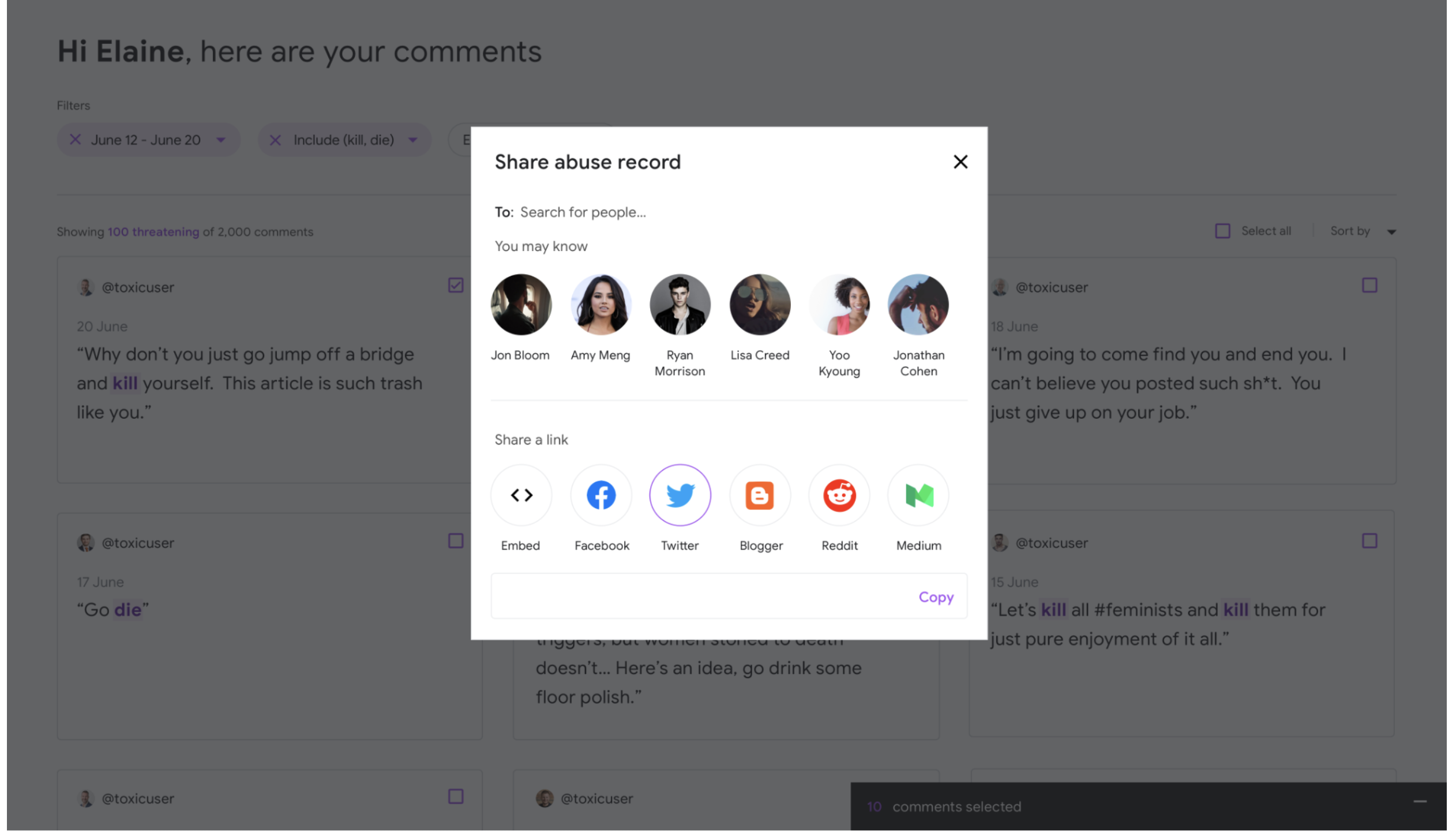}
\caption{Last page giving an opportunity to share the report with the support network }
    \label{fig:tool6}
     \hspace{-4mm}
  \end{minipage}
  \hfill
      \Description{Share the report with support network}
\end{figure*}

To instantiate these design goals, we designed a research prototype tool, as shown in Figure \ref{fig:tool3}, \ref{fig:tool4},\ref{fig:tool1},\ref{fig:tool2} and \ref{fig:tool6}. Additional pictures are available in the appendix. The prototype has the following features:
\begin{enumerate}
  \item Sign-in directly with the Platforms:  Users are required to log-in by authenticating via the platform directly, and then can only view their own data. They can not use this tool, hence, to harass someone else. This should satisfy DG1. 
  \item Aggregate Data from Platform:  The designed tool connects with the social media platform's APIs and gathers meta-data about comments (author, date/time, location, hashtags etc.) to ensure the validity of the data is maintained at the time of fetching the data. Through the API the tool initially fetches a pre-determined set of data (last 99 comments has been hard set arbitrarily to provide data worth 11 view-ports of 9 comments each) to get the user started when they log-in. This way users can view what has been happening recently in their feed. Users can continue to fetch more data as they infinitely scroll down the screen, in bursts of 27 comments with each new fetch. The data is supposed to not be stored permanently - but only kept in the cache while the user engages with the prototype to reduce any extra copies of the data. The cache is cleared when the user exits the prototype. One of the limitations of this approach is that if a user generated content has been deleted either by the platform or one of the users - a subsequent fetch will not be able to get that data. However, if the content has been removed after adding it to a Report or the fetch itself, then the deletion is inconsequential as the meta-data and the data itself has already been captured into a tangible artifact. This satisfies DG2 partially.
  \item Filters:  To satisfy DG3, the prototype enables performing retrospective analysis as well as real-time analysis for evidence collection based on filters like keywords, hashtags, names of users, and dates. As the users create new requests using these filters - the data shown on the screen changes based on the selected filters. These filters are applied to data that has already been fetched - except for the dates filter. Setting the dates filter fetches new data. 
  \item View-Restrictions: The prototype has additional filters that helps one to ignore certain data (eg. from trusted commenters, or with certain keywords) to reduce volume of data to analyze, in line with DG4.
  \item Shareable Reports: One can document the harassment by selecting one or more objectionable comments into reports and then share them in multiple ways via email, as a PDF or via social media to ensure further help is available, as expected from DG5.
 \end{enumerate}
 
The prototype was a medium-fidelity prototype created in InVision where different screens were mocked and could be clicked upon. Clicking on different pieces of the screen would enable the users to interact with the interactive elements listed above. For each user, we captured the last 99 comments prior to the session and mocked them into the screens where they can view latest comments. All the comments could be clicked to be added to a report. We also gathered particular comments from a recent harassment episode for each of the user, and these could be accessed by selecting from pre-determined dates and filter values so that the users could see how their harassment related comments might look like from a particular time or context. The users were provided a overview of how the prototype was imagined to work; and how the data was captured, saved and could be operated upon.

In the next section, we will discuss feedback from users when they were asked to use the tool to retrospectively analyze a harassment they had faced and think aloud as they used the tool.

\section{Validation phase findings}
To test how well this prototype met user needs and expectations, we invited the 18 participants from the previous phase, encouraged them to use the tool and think aloud as they tried to document and report an ongoing harassment or a harassment that they might have faced over the last year. The entire session was video recorded, after receiving participants’ informed consent. They were advised that they can stop the session at any point of time, if/when it became mentally exhausting to look at the data or otherwise, or simply leave. The recording was transcribed, coded, and themes were analyzed. Our findings are shared below



\subsection{Perceived utility}
\subsubsection{Aggregating data for documentation}

\begin{quote}
"Being able to sort of easily aggregate it all at once without having to screenshot, save, screenshot, save, all that kind of stuff, would be compelling. [Also] to be able to track it over time. I assume even if this stuff is pulled off of the platform because it's reported and seen as toxic, that this system would keep it.'' -P5
\end{quote}


Participants thought that one of the main benefits of the prototype tool was the ability to aggregate data to document online harassment attacks. As described previously, participants faced many documentation and reporting challenges. This included gathering data to service as evidence to build a case. This has also been shown to be a gap in existing patchwork of literature and existing tools, as referenced in the literature about how evidence is needed by law to act. Further, participants mentioned that screenshots were commonly used to document attacks; however, capturing and keeping track of screenshots can become a difficult task as an attack quickly spirals, as described by P5 above. One of the participants, P11, further elaborated on how taking screenshots is not something people can always do or remember. If you do not to take a screenshot (for whatever reasons) and then the content is deleted - evidence is lost forever. So, the utility of screenshots is mediated by remembering and then acting to take a screenshot above - assuming that the pile on is not happening actively where it is hard to take a screenshot every second, as mentioned by P5 above.


\subsubsection{Building a report - solving need for evidence }

\begin{quote}
``That's actually part of the reason you need a tool like this. Because if you're successfully reporting, you lose the evidence that it happened to you, if you didn't take a screenshot….Most of the journalists would benefit enormously, if they could show that there are things that were coordinated, but they would still benefit a great deal from being able to have one place to go that has a big stack of all the stuff that they're getting, to be able to print that, and show it to police, or to management, or to colleagues.'' - P11
\end{quote}

Documenting data regarding an online harassment attack is a critical part of building a case to report to platforms or law enforcement. One of the major documenting and reporting challenges is the burden of proof, which involves presenting evidence and crafting a narrative about an attack. One of the participants, P2, shared her recent experience in winning her case with the support from her lawyer to build a case based on the argument that attacking a journalist’s credibility violates her right to free speech as a journalist. Online harassment attacks can be complex when they coordinated, as mentioned in previous literature about how female journalists suffer from coordinated attacks or how they may develop across platforms. P11 described the added benefit provided by the tools ability to consolidate data to provide evidence that an attack was coordinated. 




\subsubsection{Muting/Blocking can not replace need to document and report}

\begin{quote}
``This is the problem with muting. I caution people about muting, because if they don't see a death threat, yes, that protects their mental health, but they don't know if someone just threatened to kill them, which is a really big problem." - P11
\end{quote}

Although many tools to block/mute other people have been designed previously, P11 pointed out that there are dangers to depending on a tool heavily, as one might be caught unaware about potential risks. This points to an existing gap in the design space that this research prototype fills and acts as another option for targets of harassment to manage their harassment. There are pros and cons to different options and it should be left upon the users to decide how to judiciously use different tools.  

\subsection{Future Design Considerations} 
Besides discussing the utility of the prototype, participants also reflected on what other needs remain unmet.  
\subsubsection{Adding context to report}
\begin{quote}
``An individual might receive hundreds of potentially thousands comments, but if they're not going to give context when it goes to platform, someone sitting in San Francisco is not gonna understand when someone's denigrating someone in Hausa. So you'll need to be able to highlight the specific insult. Something that's not offensive in the U.S. can be very offensive somewhere else'' - P13
\end{quote}

The prototype did not include a feature for adding notes to reports that are generated using the tool. Participants expressed interest in this feature, as current reporting mechanisms have limited opportunities to add important context to support online harassment reports. For instance, P13 pointed out that language and cultural context about an attack can be lost if content moderators had received US-centric training. This is further important for providing context to showing how this one particular incident is not a one-off incident but a pattern of continued harassment.

\subsubsection{Managing well-being while managing harassment}
\begin{quote}
``I've read online harassment about myself that I cannot forget. I can't unsee it or unhear it or unread it. Do you know what I mean? That's not something I wish to do to myself again...I want all the information there, but there's a piece of me that also just wants this content blurred on the screen unless you  scroll over it or click on it, you know I mean, unless you do something'' - P5
\end{quote}

Experiencing online harassment can be traumatic and stressful. Despite the advantages of consolidating and documenting an attack, the process of building a case typically involves reviewing the toxic content. For this reason, viewing the content from our harassment manager could be overwhelming and trigger anxiety. Participants shared design suggestions to alleviate potential emotional burdens that could be triggered from being exposed to online harassment content. A few participants suggested blurring the online harassment content. Several participants also suggested adding a pop up message as an additional buffer to warn and mentally prepare them to face the online harassment content. One of the other ways could be finding some one else to help support. 





\subsubsection{Trade Offs with Blocking/Muting when managing harassment}
\begin{quote}
``You don't also want to live in a bubble, and censor anything that's considered criticism. Because as a journalist, it's important to kind of know what people are saying. And people are gonna say things that is maybe not flattering all the time. But it could be a valid, true critique. So I don't want to, like, mute or delete, or kind of block every account that criticizes me, because sometimes they have a point. And I don't want to live in a world where everyone is just positive. That's very dangerous, information wise.'' - P4
\end{quote}

Deciding whether or not to view toxic content includes many trade-offs. While participants shared emotional and traumatic responses to viewing the contents of an online harassment attack, participants also acknowledged that choosing not to view or mute all toxic contents risks can trap them in a filter bubble or miss a real threat.

\subsubsection{Managing harassment across multiple platforms}
\begin{quote}
``I think that’s another thing too that harassment comes from different sides. It’s coming from phones, text messages, email, a variety of social media platforms. It’s sometimes coming from the news media itself, who might be kind of like aggrandizing particular things or parroting you.'' - P7
\end{quote}

The interviews revealed that online harassment attacks can occur across several platforms and channels. Several participants described receiving harassment through various avenues: social media platforms, texts, emails, and comment sections under articles and content on publisher sites. Cross-platform online harassment can transpire quickly within short or slowly throughout longer time spans. For instance, online harassment attacks can achieve virality, as journalists’ articles or posts published online gain traction and engagement. For instance, P7’s journalistic activities and controversial affiliation with a media publication sparked a series of online harassment on a platform, in comment sections where her articles were published, and a media outlet’s video channel. Cross-platform harassment adds complexities to challenges in understanding, documenting, and reporting attacks because now the targets needs to cross-examine across multiple surfaces with differential levels of access, and support without having sophisticated data science tools.

\section{Discussion}
Our results indicate the reality of under-prepared journalists and activists managing  crisis of attack and trying to recover from these attacks alone with limited success owing to the large scale of such attacks, and lack of appropriate tools. While multiple tools have been designed that enable one to filter content (Blockparty \cite{BlockParty}) or block/mute content (features provided by platforms), participants have shown that such temporary fixes can have large scale repercussions where one might miss out life threatening threats - a reality that does happen to journalists from India and Malta \cite{posetti2018violence}. Similarly, while tools that provide real-time support through access to resources (Hollaback \cite{blackwell2017classification}, SquadBox \cite{mahar2018squadbox}) or emotional support by sending positive vibes (Troll-Busters \cite{ferrier2018trollbusters}) are incredibly useful, our participants do need additional support in real-time to gather evidence. Recent tools like Unmochon \cite{sultana2021unmochon} are an example of servicing this need using screenshots - however, screenshots based on protracted relationships with admins on Facebook groups are not equitably accessible to all, and might not work for those under high velocity attacks where taking screenshots is impossible. We found that we need to move to an approach that continues to empower journalists/activists to document and create reports of their harassment - while focusing on privacy considerations, capturing valid data, enabling analysis without sophisticated tools, managing volume by filtering out the non-relevant, and creating shareable artefacts like reports that can be actioned upon. We identify opportunities to further expand HCI’s role as the "friend" of targets of harassment and its commitment to designing to make a societal difference. 

While our work has focused on journalists and activists, we believe that through such designs HCI can be generalized to enabling targets of harassment that belong to other groups and minorities too. Folks like teachers and county clerks, as shown in works by \citet{woodruff2014necessary}, also feel disempowered in managing their reputation online and have shown to have a desire to fix this societal problem \citet{kuzminykh2016people}. Such folks who receive harassment online and have limited opportunities enact upon them to fix this problem have similar needs to be able to document the incoming harassment, create reports about them and then share these reports with people who are more powerful or better resourced to help the targets of harassment.

\subsection{PMCR Framework and drawing attention to PM vs. CR needs}
The chronological and PMCR framework based analysis revealed that needs of targets of harassment were not limited to a single stage in a targets’ harassment trajectory, as participants referenced several PMCR needs that spanned across the chronological life cycle of an online harassment attack. Viewing from the perspective of PMCR \cite{jigsawmedium} 
- we have shown the value of applying this existing framework that can be used to understand and classify user needs based on the complex trajectories of harassment as opposed to a simplistic chronological trajectory of before, during and after harassment. We believe that this is a contribution, providing an example for designers of applying the PMCR framework to dig deeper and explore the design space for P, M, C, R, and at the intersections of these needs as well. In this paper we focused on the intersection of CR. Hence much more needs to be done still.

Currently, there is a heavy focus on fulfilling targets’ PMCR needs through scaling content moderation using automated techniques. Our literature review shows that many technical solutions are centered around scaling automated content moderation efforts that are implemented at the platform level \cite{jhaver2018online, mathew2019thou, scheuerman2018safe}. Platforms have employed content moderation strategies that are geared towards prevention and monitoring needs, by preventing the most toxic forms of content from being published online and surveilling potential threats by using automated detection techniques. While there have been great advances in automated detection strategies, executing text-based machine learning and natural language processing techniques at scale has limitations due to challenges accessing and creating robust training datasets that capture nuances in context, tone, and cultures \cite{scheuerman2018safe, jhaver2018online}. Since targets’ crisis and recovery needs are mainly addressed through reporting mechanisms, like when a target or user reports a post as inappropriate, an immense burden is shifted onto targets across all the stages of harassment. The presented prototype, albeit provides a streamlined process, still lays burden of management of harassment on the targets. This is the first step in the wide design space and we point to a new research direction where we should consider focusing on automated help during crisis and recovery, and not just prevention or monitoring stages of harassment. This research direction needs to explore privacy concerns across PMCR, and how well current AI tools may really perform ? Similarly, policy makers now have an opportunity to decide if the onus of this emotional and physical labor should fall upon the targets during CR or the platforms during PM.

There is an even wider space on how to help every Internet user and not just journalists and activists to navigate harassment lobbied at them actively or passively as they generate and consume content online. This includes design spaces within Prevention/Monitoring like identifying that they are under attack, or what an attack even means/constitutes ? Since journalists/activists have been under attack for so long, they have started to learn answers to these questions. But, those of us who are lesser experienced - we need even more resources to learn and design about this space.

\subsection{Drawing attention to designing for mental well-being}
Despite being able to use a tool to document and report harassment, participants in this research showed that this is an incredibly hard space to design for. This is a global issue that multiple researchers are already working on in many different ways, like building tools  \cite{mahar2018squadbox, molina2019harassment, wijesiriwardene2020alone, arora2020novel, mishra2019tackling, lowd2018can, kwon2018tweety}; understanding cyberstalking \cite{frommholz2016textual, al2016technology}, cyber bullying \cite{van2018automatic}, incivility \cite{davidson2020developing} and improving content moderation \cite{seering2020reconsidering, perspectiveapi} and highlighting impact of Russian trolls on harassment \cite{im2020still}. However, limited  work is happening to push this direction further, as shown by \citet{sultana2021unmochon} who reiterated the significance in considering methods for authenticating documented evidence for online harassment attacks. We take inspiration from this previous work and contribute to opening up multiple new research and design directions for designers and researchers of online harassment design space to consider the mental well-being of the targets deeply as they are mostly under prepared, and alone while trying to deal with attacks for doing their profession. Another direction is how do we reduce this loneliness that haunts the targets, by helping them find support networks - especially when they need them the most! These support networks can be friends, family members, NGO/NFPs, law enforcement, lawyers, platforms, etc. Further, how can we empower this support network directly to reduce this burden on the targets of harassment and protect their well-being ? Members of these support networks can also benefit from tools that can help them make sense of this data \cite{GIS2013, GIS2014} to create unbiased evidence \cite{GST2016, GCS2017} . These directions are in conjunction with the four design goals that we used to guide our design: Privacy Considerations about who can access a target's data, supporting Data Validity of the evidence collected, enabling Data-Analysis by targets without sophisticated tools, providing support to Manage Volume of harassment, and creating artifacts that can be Shared with Support Networks.

\subsection{Inaction is not neutral - it is a societal risk}
Participants shared that being able to document online harassment incidents validates their experiences and provides a sense of relief that the evidence exists, especially as content online is easily deleted. Facilitating data aggregation allows targets to document an online harassment attack while minimizing the emotional and laborious burden of collecting evidence and cleaning up the aftermath, as journalists often delete content containing online harassment from their social media accounts \cite{ringel2020proactive}. As platforms, researchers, and advocacy organizations have invested in building tools to combat online harassment \cite{mahar2018squadbox,ferrier2018trollbusters, blackwell2017classification, sultana2021unmochon, kwon2018tweety, BlockParty, perspectiveapi, lowd2018}, targets of online harassment have also practiced self-driven strategies, mostly involving avoidance and self-censorship on social media, to minimize exposure to online harassment and its associated threats \cite{posetti2020online, arora2020novel, holton2021not}. UNESCO and ICJ’s global survey revealed that women reported the following as common self-driven strategies: making themselves less public online, self-censoring on social media, withdrawing from all online interactions, avoiding engagement with their audience \cite{posetti2020online}. Discussions throughout our study also frequently mentioned that participants have avoided and self-censored on social media after experiencing severe online harassment attacks out of fear and as safety precautions. If, we as a community and larger society do not prioritize these research directions further, we stand risk of loosing voices of many people online - especially those who belong to minority groups and further have chosen to spend their time investigating facts and sharing them widely, like the ones from Malta and India. This space needs more resources, from technology perspective to manage the volume of harassment, communication perspective on how to encourage conducive online conversations, and humanistic perspective to highlight that this is a societal challenge. This is a challenge not just for the journalists and activists but pretty much anyone who can be made a target.



\section{Limitations}
Our work had several limitations. First, convenience and snowball sampling was used to recruit participants for the focus group and interviews, so the participants were found through the researchers’ first and second degree connections, which contributed to minor sampling biases. Most of the participants were located in the US. Additionally, the same group of participants were included in the phase one and three interviews to work around recruitment challenges and save time for building trust and repertoire in discussions that could involve sensitive and traumatic online harassment experiences. Second, we focused the scope on a small part of online harassment and the problem space is much larger. Third, this work primarily focused on mixed-methods qualitative research of focus group and interviews. Future work should consider alternative approaches to understand this space further.

\section{Conclusion}
In this paper, we conducted focus groups and interviews to learn about female journalists' experiences with online harassment and gathered feedback on a prototype of a tool that facilitates documentation and reporting online harassment attacks on social media platforms. Based on our findings, we applied the PMCR framework \cite{jigsawmedium} 
as a way of categorizing key types of needs throughout the different stages of online harassment, showing that targets need help to Prevent, Monitor, manage a Crisis, and Recover from crisis. As existing technical solutions focus heavily on prevention and monitoring using machine learning and content moderation practice, we designed a prototype focusing on crisis and recovery needs, specifically documentation and reporting. Tackling documentation and reporting challenges is an important effort for empowering female journalists and their support networks to address online harassment attacks and to progress towards equity in the digital public sphere.

\section{Acknowledgements}
We would like to acknowledge the work done by multiple teams at Google who developed and evolved the PMCR framework over multiple years (including but not limited to teams working with at-risk users at Google and Jigsaw).


\bibliographystyle{ACM-Reference-Format}
\bibliography{main.bib}


\begin{thebibliography}{82}


\ifx \showCODEN    \undefined \def \showCODEN     #1{\unskip}     \fi
\ifx \showDOI      \undefined \def \showDOI       #1{#1}\fi
\ifx \showISBNx    \undefined \def \showISBNx     #1{\unskip}     \fi
\ifx \showISBNxiii \undefined \def \showISBNxiii  #1{\unskip}     \fi
\ifx \showISSN     \undefined \def \showISSN      #1{\unskip}     \fi
\ifx \showLCCN     \undefined \def \showLCCN      #1{\unskip}     \fi
\ifx \shownote     \undefined \def \shownote      #1{#1}          \fi
\ifx \showarticletitle \undefined \def \showarticletitle #1{#1}   \fi
\ifx \showURL      \undefined \def \showURL       {\relax}        \fi
\providecommand\bibfield[2]{#2}
\providecommand\bibinfo[2]{#2}
\providecommand\natexlab[1]{#1}
\providecommand\showeprint[2][]{arXiv:#2}

\bibitem[\protect\citeauthoryear{al~Khateeb and Epiphaniou}{al~Khateeb and Epiphaniou}{2016}]%
        {al2016technology}
\bibfield{author}{\bibinfo{person}{Haider~M al Khateeb} {and} \bibinfo{person}{Gregory Epiphaniou}.} \bibinfo{year}{2016}\natexlab{}.
\newblock \showarticletitle{How technology can mitigate and counteract cyber-stalking and online grooming}.
\newblock \bibinfo{journal}{\emph{Computer Fraud \& Security}} \bibinfo{volume}{2016}, \bibinfo{number}{1} (\bibinfo{year}{2016}), \bibinfo{pages}{14--18}.
\newblock


\bibitem[\protect\citeauthoryear{Andalibi, Haimson, De~Choudhury, and Forte}{Andalibi et~al\mbox{.}}{2016}]%
        {andalibi2016understanding}
\bibfield{author}{\bibinfo{person}{Nazanin Andalibi}, \bibinfo{person}{Oliver~L Haimson}, \bibinfo{person}{Munmun De~Choudhury}, {and} \bibinfo{person}{Andrea Forte}.} \bibinfo{year}{2016}\natexlab{}.
\newblock \showarticletitle{Understanding social media disclosures of sexual abuse through the lenses of support seeking and anonymity}. In \bibinfo{booktitle}{\emph{Proceedings of the 2016 CHI conference on human factors in computing systems}}. \bibinfo{pages}{3906--3918}.
\newblock


\bibitem[\protect\citeauthoryear{Antunovic}{Antunovic}{2019}]%
        {antunovic2019we}
\bibfield{author}{\bibinfo{person}{Dunja Antunovic}.} \bibinfo{year}{2019}\natexlab{}.
\newblock \showarticletitle{“We wouldn’t say it to their faces”: Online harassment, women sports journalists, and feminism}.
\newblock \bibinfo{journal}{\emph{Feminist Media Studies}} \bibinfo{volume}{19}, \bibinfo{number}{3} (\bibinfo{year}{2019}), \bibinfo{pages}{428--442}.
\newblock


\bibitem[\protect\citeauthoryear{Arora, Guo, Levitan, McGregor, and Hirschberg}{Arora et~al\mbox{.}}{2020}]%
        {arora2020novel}
\bibfield{author}{\bibinfo{person}{Ishaan Arora}, \bibinfo{person}{Julia Guo}, \bibinfo{person}{Sarah~Ita Levitan}, \bibinfo{person}{Susan McGregor}, {and} \bibinfo{person}{Julia Hirschberg}.} \bibinfo{year}{2020}\natexlab{}.
\newblock \showarticletitle{A novel methodology for developing automatic harassment classifiers for Twitter}. In \bibinfo{booktitle}{\emph{Proceedings of the Fourth Workshop on Online Abuse and Harms}}. \bibinfo{pages}{7--15}.
\newblock


\bibitem[\protect\citeauthoryear{Ashktorab and Vitak}{Ashktorab and Vitak}{2016}]%
        {ashktorab2016designing}
\bibfield{author}{\bibinfo{person}{Zahra Ashktorab} {and} \bibinfo{person}{Jessica Vitak}.} \bibinfo{year}{2016}\natexlab{}.
\newblock \showarticletitle{Designing cyberbullying mitigation and prevention solutions through participatory design with teenagers}. In \bibinfo{booktitle}{\emph{Proceedings of the 2016 CHI Conference on Human Factors in Computing Systems}}. \bibinfo{pages}{3895--3905}.
\newblock


\bibitem[\protect\citeauthoryear{Backe, Lilleston, and McCleary-Sills}{Backe et~al\mbox{.}}{2018}]%
        {backe2018networked}
\bibfield{author}{\bibinfo{person}{Emma~Louise Backe}, \bibinfo{person}{Pamela Lilleston}, {and} \bibinfo{person}{Jennifer McCleary-Sills}.} \bibinfo{year}{2018}\natexlab{}.
\newblock \showarticletitle{Networked individuals, gendered violence: A literature review of cyberviolence}.
\newblock \bibinfo{journal}{\emph{Violence and gender}} \bibinfo{volume}{5}, \bibinfo{number}{3} (\bibinfo{year}{2018}), \bibinfo{pages}{135--146}.
\newblock


\bibitem[\protect\citeauthoryear{Belair-Gagnon, Nelson, and Lewis}{Belair-Gagnon et~al\mbox{.}}{2019}]%
        {belair2019audience}
\bibfield{author}{\bibinfo{person}{Valerie Belair-Gagnon}, \bibinfo{person}{Jacob~L Nelson}, {and} \bibinfo{person}{Seth~C Lewis}.} \bibinfo{year}{2019}\natexlab{}.
\newblock \showarticletitle{Audience engagement, reciprocity, and the pursuit of community connectedness in public media journalism}.
\newblock \bibinfo{journal}{\emph{Journalism Practice}} \bibinfo{volume}{13}, \bibinfo{number}{5} (\bibinfo{year}{2019}), \bibinfo{pages}{558--575}.
\newblock


\bibitem[\protect\citeauthoryear{Blackwell, Chen, Schoenebeck, and Lampe}{Blackwell et~al\mbox{.}}{2018}]%
        {blackwell2018online}
\bibfield{author}{\bibinfo{person}{Lindsay Blackwell}, \bibinfo{person}{Tianying Chen}, \bibinfo{person}{Sarita Schoenebeck}, {and} \bibinfo{person}{Cliff Lampe}.} \bibinfo{year}{2018}\natexlab{}.
\newblock \showarticletitle{When online harassment is perceived as justified}. In \bibinfo{booktitle}{\emph{Proceedings of the International AAAI Conference on Web and Social Media}}, Vol.~\bibinfo{volume}{12}.
\newblock


\bibitem[\protect\citeauthoryear{Blackwell, Dimond, Schoenebeck, and Lampe}{Blackwell et~al\mbox{.}}{2017}]%
        {blackwell2017classification}
\bibfield{author}{\bibinfo{person}{Lindsay Blackwell}, \bibinfo{person}{Jill Dimond}, \bibinfo{person}{Sarita Schoenebeck}, {and} \bibinfo{person}{Cliff Lampe}.} \bibinfo{year}{2017}\natexlab{}.
\newblock \showarticletitle{Classification and its consequences for online harassment: Design insights from heartmob}.
\newblock \bibinfo{journal}{\emph{Proceedings of the ACM on Human-Computer Interaction}} \bibinfo{volume}{1}, \bibinfo{number}{CSCW} (\bibinfo{year}{2017}), \bibinfo{pages}{1--19}.
\newblock


\bibitem[\protect\citeauthoryear{by~Hollaback!}{by~Hollaback!}{nd}]%
        {heartmob}
\bibfield{author}{\bibinfo{person}{Heartmob by Hollaback!}} \bibinfo{year}{n.d.}\natexlab{}.
\newblock \bibinfo{title}{Know Your Rights!}
\newblock
\newblock
\urldef\tempurl%
\url{https://iheartmob.org/resources/rights}
\showURL{%
Retrieved September 9, 2021 from \tempurl}


\bibitem[\protect\citeauthoryear{Carlson and Witt}{Carlson and Witt}{2020}]%
        {carlson2020online}
\bibfield{author}{\bibinfo{person}{Caitlin~Ring Carlson} {and} \bibinfo{person}{Haley Witt}.} \bibinfo{year}{2020}\natexlab{}.
\newblock \showarticletitle{Online harassment of US women journalists and its impact on press freedom}.
\newblock \bibinfo{journal}{\emph{First Monday}} (\bibinfo{year}{2020}).
\newblock


\bibitem[\protect\citeauthoryear{Chadha, Steiner, Vitak, and Ashktorab}{Chadha et~al\mbox{.}}{2020}]%
        {chadha2020women}
\bibfield{author}{\bibinfo{person}{Kalyani Chadha}, \bibinfo{person}{Linda Steiner}, \bibinfo{person}{Jessica Vitak}, {and} \bibinfo{person}{Zahra Ashktorab}.} \bibinfo{year}{2020}\natexlab{}.
\newblock \showarticletitle{Women’s responses to online harassment}.
\newblock \bibinfo{journal}{\emph{International Journal of Communication}}  \bibinfo{volume}{14} (\bibinfo{year}{2020}), \bibinfo{pages}{19}.
\newblock


\bibitem[\protect\citeauthoryear{Chandaluri and Phadke}{Chandaluri and Phadke}{2019}]%
        {chandaluri2019cross}
\bibfield{author}{\bibinfo{person}{Rohit~Kumar Chandaluri} {and} \bibinfo{person}{Shruti Phadke}.} \bibinfo{year}{2019}\natexlab{}.
\newblock \emph{\bibinfo{title}{Cross-Platform Data Collection and Analysis for Online Hate Groups}}.
\newblock \bibinfo{thesistype}{Ph.\,D. Dissertation}. \bibinfo{school}{Virginia Tech}.
\newblock


\bibitem[\protect\citeauthoryear{Chen, Pain, Chen, Mekelburg, Springer, and Troger}{Chen et~al\mbox{.}}{2020}]%
        {chen2020you}
\bibfield{author}{\bibinfo{person}{Gina~Masullo Chen}, \bibinfo{person}{Paromita Pain}, \bibinfo{person}{Victoria~Y Chen}, \bibinfo{person}{Madlin Mekelburg}, \bibinfo{person}{Nina Springer}, {and} \bibinfo{person}{Franziska Troger}.} \bibinfo{year}{2020}\natexlab{}.
\newblock \showarticletitle{‘You really have to have a thick skin’: A cross-cultural perspective on how online harassment influences female journalists}.
\newblock \bibinfo{journal}{\emph{Journalism}} \bibinfo{volume}{21}, \bibinfo{number}{7} (\bibinfo{year}{2020}), \bibinfo{pages}{877--895}.
\newblock


\bibitem[\protect\citeauthoryear{Chen, Pain, and Zhang}{Chen et~al\mbox{.}}{2018}]%
        {chen2018nastywomen}
\bibfield{author}{\bibinfo{person}{Gina~Masullo Chen}, \bibinfo{person}{Paromita Pain}, {and} \bibinfo{person}{Jinglun Zhang}.} \bibinfo{year}{2018}\natexlab{}.
\newblock \showarticletitle{\# NastyWomen: Reclaiming the Twitterverse from misogyny}.
\newblock In \bibinfo{booktitle}{\emph{Mediating misogyny}}. \bibinfo{publisher}{Springer}, \bibinfo{pages}{371--388}.
\newblock


\bibitem[\protect\citeauthoryear{Citron}{Citron}{2014}]%
        {citron2014addressing}
\bibfield{author}{\bibinfo{person}{Danielle~Keats Citron}.} \bibinfo{year}{2014}\natexlab{}.
\newblock \showarticletitle{Addressing cyber harassment: An overview of hate crimes in cyberspace}.
\newblock \bibinfo{journal}{\emph{Case W. Res. JL Tech. \& Internet}}  \bibinfo{volume}{6} (\bibinfo{year}{2014}), \bibinfo{pages}{1}.
\newblock


\bibitem[\protect\citeauthoryear{Davidson, Sun, and Wojcieszak}{Davidson et~al\mbox{.}}{2020}]%
        {davidson2020developing}
\bibfield{author}{\bibinfo{person}{Sam Davidson}, \bibinfo{person}{Qiusi Sun}, {and} \bibinfo{person}{Magdalena Wojcieszak}.} \bibinfo{year}{2020}\natexlab{}.
\newblock \showarticletitle{Developing a new classifier for automated identification of incivility in social media}. In \bibinfo{booktitle}{\emph{Proceedings of the fourth workshop on online abuse and harms}}. \bibinfo{pages}{95--101}.
\newblock


\bibitem[\protect\citeauthoryear{Duggan}{Duggan}{2017}]%
        {duggan2017online}
\bibfield{author}{\bibinfo{person}{Maeve Duggan}.} \bibinfo{year}{2017}\natexlab{}.
\newblock \showarticletitle{Online harassment 2017}.
\newblock  (\bibinfo{year}{2017}).
\newblock


\bibitem[\protect\citeauthoryear{Ferrier and Garud-Patkar}{Ferrier and Garud-Patkar}{2018}]%
        {ferrier2018trollbusters}
\bibfield{author}{\bibinfo{person}{Michelle Ferrier} {and} \bibinfo{person}{Nisha Garud-Patkar}.} \bibinfo{year}{2018}\natexlab{}.
\newblock \showarticletitle{TrollBusters: Fighting online harassment of women journalists}.
\newblock In \bibinfo{booktitle}{\emph{Mediating Misogyny}}. \bibinfo{publisher}{Springer}, \bibinfo{pages}{311--332}.
\newblock


\bibitem[\protect\citeauthoryear{Fox and Tang}{Fox and Tang}{2017}]%
        {fox2017women}
\bibfield{author}{\bibinfo{person}{Jesse Fox} {and} \bibinfo{person}{Wai~Yen Tang}.} \bibinfo{year}{2017}\natexlab{}.
\newblock \showarticletitle{Women’s experiences with general and sexual harassment in online video games: Rumination, organizational responsiveness, withdrawal, and coping strategies}.
\newblock \bibinfo{journal}{\emph{New media \& society}} \bibinfo{volume}{19}, \bibinfo{number}{8} (\bibinfo{year}{2017}), \bibinfo{pages}{1290--1307}.
\newblock


\bibitem[\protect\citeauthoryear{Frommholz, Al-Khateeb, Potthast, Ghasem, Shukla, and Short}{Frommholz et~al\mbox{.}}{2016}]%
        {frommholz2016textual}
\bibfield{author}{\bibinfo{person}{Ingo Frommholz}, \bibinfo{person}{Haider~M Al-Khateeb}, \bibinfo{person}{Martin Potthast}, \bibinfo{person}{Zinnar Ghasem}, \bibinfo{person}{Mitul Shukla}, {and} \bibinfo{person}{Emma Short}.} \bibinfo{year}{2016}\natexlab{}.
\newblock \showarticletitle{On textual analysis and machine learning for cyberstalking detection}.
\newblock \bibinfo{journal}{\emph{Datenbank-Spektrum}} \bibinfo{volume}{16}, \bibinfo{number}{2} (\bibinfo{year}{2016}), \bibinfo{pages}{127--135}.
\newblock


\bibitem[\protect\citeauthoryear{Gillespie}{Gillespie}{2020}]%
        {gillespie2020content}
\bibfield{author}{\bibinfo{person}{Tarleton Gillespie}.} \bibinfo{year}{2020}\natexlab{}.
\newblock \showarticletitle{Content moderation, AI, and the question of scale}.
\newblock \bibinfo{journal}{\emph{Big Data \& Society}} \bibinfo{volume}{7}, \bibinfo{number}{2} (\bibinfo{year}{2020}), \bibinfo{pages}{2053951720943234}.
\newblock


\bibitem[\protect\citeauthoryear{Google}{Google}{nd}]%
        {perspectiveapi}
\bibfield{author}{\bibinfo{person}{Google}.} \bibinfo{year}{n.d.}\natexlab{}.
\newblock \bibinfo{title}{Perspective API}.
\newblock
\newblock
\urldef\tempurl%
\url{https://perspectiveapi.com/}
\showURL{%
Retrieved September 9, 2021 from \tempurl}


\bibitem[\protect\citeauthoryear{Gorwa, Binns, and Katzenbach}{Gorwa et~al\mbox{.}}{2020}]%
        {gorwa2020algorithmic}
\bibfield{author}{\bibinfo{person}{Robert Gorwa}, \bibinfo{person}{Reuben Binns}, {and} \bibinfo{person}{Christian Katzenbach}.} \bibinfo{year}{2020}\natexlab{}.
\newblock \showarticletitle{Algorithmic content moderation: Technical and political challenges in the automation of platform governance}.
\newblock \bibinfo{journal}{\emph{Big Data \& Society}} \bibinfo{volume}{7}, \bibinfo{number}{1} (\bibinfo{year}{2020}), \bibinfo{pages}{2053951719897945}.
\newblock


\bibitem[\protect\citeauthoryear{Goyal and Fussell}{Goyal and Fussell}{2016}]%
        {GST2016}
\bibfield{author}{\bibinfo{person}{Nitesh Goyal} {and} \bibinfo{person}{Susan~R. Fussell}.} \bibinfo{year}{2016}\natexlab{}.
\newblock \showarticletitle{Effects of Sensemaking Translucence on Distributed Collaborative Analysis}. In \bibinfo{booktitle}{\emph{Proceedings of the 19th ACM Conference on Computer-Supported Cooperative Work \& Social Computing}} (San Francisco, California, USA) \emph{(\bibinfo{series}{CSCW '16})}. \bibinfo{publisher}{Association for Computing Machinery}, \bibinfo{address}{New York, NY, USA}, \bibinfo{pages}{288–302}.
\newblock
\showISBNx{9781450335928}
\urldef\tempurl%
\url{https://doi.org/10.1145/2818048.2820071}
\showDOI{\tempurl}


\bibitem[\protect\citeauthoryear{Goyal and Fussell}{Goyal and Fussell}{2017}]%
        {GCS2017}
\bibfield{author}{\bibinfo{person}{Nitesh Goyal} {and} \bibinfo{person}{Susan~R. Fussell}.} \bibinfo{year}{2017}\natexlab{}.
\newblock \showarticletitle{Intelligent Interruption Management Using Electro Dermal Activity Based Physiological Sensor for Collaborative Sensemaking}.
\newblock \bibinfo{journal}{\emph{Proc. ACM Interact. Mob. Wearable Ubiquitous Technol.}} \bibinfo{volume}{1}, \bibinfo{number}{3}, Article \bibinfo{articleno}{52} (\bibinfo{date}{sep} \bibinfo{year}{2017}), \bibinfo{numpages}{21}~pages.
\newblock
\urldef\tempurl%
\url{https://doi.org/10.1145/3130917}
\showDOI{\tempurl}


\bibitem[\protect\citeauthoryear{Goyal, Leshed, Cosley, and Fussell}{Goyal et~al\mbox{.}}{2014}]%
        {GIS2014}
\bibfield{author}{\bibinfo{person}{Nitesh Goyal}, \bibinfo{person}{Gilly Leshed}, \bibinfo{person}{Dan Cosley}, {and} \bibinfo{person}{Susan~R. Fussell}.} \bibinfo{year}{2014}\natexlab{}.
\newblock \showarticletitle{Effects of Implicit Sharing in Collaborative Analysis}. In \bibinfo{booktitle}{\emph{Proceedings of the SIGCHI Conference on Human Factors in Computing Systems}} (Toronto, Ontario, Canada) \emph{(\bibinfo{series}{CHI '14})}. \bibinfo{publisher}{Association for Computing Machinery}, \bibinfo{address}{New York, NY, USA}, \bibinfo{pages}{129–138}.
\newblock
\showISBNx{9781450324731}
\urldef\tempurl%
\url{https://doi.org/10.1145/2556288.2557229}
\showDOI{\tempurl}


\bibitem[\protect\citeauthoryear{Goyal, Leshed, and Fussell}{Goyal et~al\mbox{.}}{2013}]%
        {GIS2013}
\bibfield{author}{\bibinfo{person}{Nitesh Goyal}, \bibinfo{person}{Gilly Leshed}, {and} \bibinfo{person}{Susan~R. Fussell}.} \bibinfo{year}{2013}\natexlab{}.
\newblock \showarticletitle{Effects of Visualization and Note-Taking on Sensemaking and Analysis}. In \bibinfo{booktitle}{\emph{Proceedings of the SIGCHI Conference on Human Factors in Computing Systems}} (Paris, France) \emph{(\bibinfo{series}{CHI '13})}. \bibinfo{publisher}{Association for Computing Machinery}, \bibinfo{address}{New York, NY, USA}, \bibinfo{pages}{2721–2724}.
\newblock
\showISBNx{9781450318990}
\urldef\tempurl%
\url{https://doi.org/10.1145/2470654.2481376}
\showDOI{\tempurl}


\bibitem[\protect\citeauthoryear{Guti{\'e}rrez-Esparza, Vallejo-Allende, and Hern{\'a}ndez-Torruco}{Guti{\'e}rrez-Esparza et~al\mbox{.}}{2019}]%
        {gutierrez2019classification}
\bibfield{author}{\bibinfo{person}{Guadalupe~Obdulia Guti{\'e}rrez-Esparza}, \bibinfo{person}{Maite Vallejo-Allende}, {and} \bibinfo{person}{Jos{\'e} Hern{\'a}ndez-Torruco}.} \bibinfo{year}{2019}\natexlab{}.
\newblock \showarticletitle{Classification of cyber-aggression cases applying machine learning}.
\newblock \bibinfo{journal}{\emph{Applied Sciences}} \bibinfo{volume}{9}, \bibinfo{number}{9} (\bibinfo{year}{2019}), \bibinfo{pages}{1828}.
\newblock


\bibitem[\protect\citeauthoryear{Hamin and Rosli}{Hamin and Rosli}{2018}]%
        {hamin2018cloaked}
\bibfield{author}{\bibinfo{person}{Zaiton Hamin} {and} \bibinfo{person}{Wan Rosalili~Wan Rosli}.} \bibinfo{year}{2018}\natexlab{}.
\newblock \showarticletitle{Cloaked by cyber space: A legal response to the risks of cyber stalking in Malaysia}.
\newblock \bibinfo{journal}{\emph{International Journal of Cyber Criminology}} \bibinfo{volume}{12}, \bibinfo{number}{1} (\bibinfo{year}{2018}), \bibinfo{pages}{316--332}.
\newblock


\bibitem[\protect\citeauthoryear{Holton, B{\'e}lair-Gagnon, Bossio, and Molyneux}{Holton et~al\mbox{.}}{2021}]%
        {holton2021not}
\bibfield{author}{\bibinfo{person}{Avery~E Holton}, \bibinfo{person}{Val{\'e}rie B{\'e}lair-Gagnon}, \bibinfo{person}{Diana Bossio}, {and} \bibinfo{person}{Logan Molyneux}.} \bibinfo{year}{2021}\natexlab{}.
\newblock \showarticletitle{“Not Their Fault, but Their Problem”: Organizational Responses to the Online Harassment of Journalists}.
\newblock \bibinfo{journal}{\emph{Journalism Practice}} (\bibinfo{year}{2021}), \bibinfo{pages}{1--16}.
\newblock


\bibitem[\protect\citeauthoryear{Im, Chandrasekharan, Sargent, Lighthammer, Denby, Bhargava, Hemphill, Jurgens, and Gilbert}{Im et~al\mbox{.}}{2020}]%
        {im2020still}
\bibfield{author}{\bibinfo{person}{Jane Im}, \bibinfo{person}{Eshwar Chandrasekharan}, \bibinfo{person}{Jackson Sargent}, \bibinfo{person}{Paige Lighthammer}, \bibinfo{person}{Taylor Denby}, \bibinfo{person}{Ankit Bhargava}, \bibinfo{person}{Libby Hemphill}, \bibinfo{person}{David Jurgens}, {and} \bibinfo{person}{Eric Gilbert}.} \bibinfo{year}{2020}\natexlab{}.
\newblock \showarticletitle{Still out there: Modeling and identifying russian troll accounts on twitter}. In \bibinfo{booktitle}{\emph{12th ACM Conference on Web Science}}. \bibinfo{pages}{1--10}.
\newblock


\bibitem[\protect\citeauthoryear{Jane}{Jane}{2017}]%
        {jane2017gendered}
\bibfield{author}{\bibinfo{person}{Emma~A Jane}.} \bibinfo{year}{2017}\natexlab{}.
\newblock \showarticletitle{Gendered cyberhate, victim-blaming, and why the internet is more like driving a car on a road than being naked in the snow}.
\newblock In \bibinfo{booktitle}{\emph{Cybercrime and its victims}}. \bibinfo{publisher}{Routledge}, \bibinfo{pages}{61--78}.
\newblock


\bibitem[\protect\citeauthoryear{Jane}{Jane}{2020}]%
        {jane2020online}
\bibfield{author}{\bibinfo{person}{Emma~A Jane}.} \bibinfo{year}{2020}\natexlab{}.
\newblock \showarticletitle{Online Abuse and Harassment}.
\newblock \bibinfo{journal}{\emph{The International Encyclopedia of Gender, Media, and Communication}} (\bibinfo{year}{2020}), \bibinfo{pages}{1--16}.
\newblock


\bibitem[\protect\citeauthoryear{Jhaver}{Jhaver}{2020}]%
        {jhaver2020identifying}
\bibfield{author}{\bibinfo{person}{Shagun Jhaver}.} \bibinfo{year}{2020}\natexlab{}.
\newblock \emph{\bibinfo{title}{Identifying opportunities to improve content moderation}}.
\newblock \bibinfo{thesistype}{Ph.\,D. Dissertation}. \bibinfo{school}{Georgia Institute of Technology}.
\newblock


\bibitem[\protect\citeauthoryear{Jhaver, Ghoshal, Bruckman, and Gilbert}{Jhaver et~al\mbox{.}}{2018}]%
        {jhaver2018online}
\bibfield{author}{\bibinfo{person}{Shagun Jhaver}, \bibinfo{person}{Sucheta Ghoshal}, \bibinfo{person}{Amy Bruckman}, {and} \bibinfo{person}{Eric Gilbert}.} \bibinfo{year}{2018}\natexlab{}.
\newblock \showarticletitle{Online harassment and content moderation: The case of blocklists}.
\newblock \bibinfo{journal}{\emph{ACM Transactions on Computer-Human Interaction (TOCHI)}} \bibinfo{volume}{25}, \bibinfo{number}{2} (\bibinfo{year}{2018}), \bibinfo{pages}{1--33}.
\newblock


\bibitem[\protect\citeauthoryear{Kuzminykh and Lank}{Kuzminykh and Lank}{2016}]%
        {kuzminykh2016people}
\bibfield{author}{\bibinfo{person}{Anastasia Kuzminykh} {and} \bibinfo{person}{Edward Lank}.} \bibinfo{year}{2016}\natexlab{}.
\newblock \showarticletitle{People searched by people: Context-based selectiveness in online search}. In \bibinfo{booktitle}{\emph{Proceedings of the 2016 ACM Conference on Designing Interactive Systems}}. \bibinfo{pages}{749--760}.
\newblock


\bibitem[\protect\citeauthoryear{Kwon, Liang, Tandon, Berman, Chang, and Gilbert}{Kwon et~al\mbox{.}}{2018}]%
        {kwon2018tweety}
\bibfield{author}{\bibinfo{person}{Saebom Kwon}, \bibinfo{person}{Puhe Liang}, \bibinfo{person}{Sonali Tandon}, \bibinfo{person}{Jacob Berman}, \bibinfo{person}{Pai-ju Chang}, {and} \bibinfo{person}{Eric Gilbert}.} \bibinfo{year}{2018}\natexlab{}.
\newblock \showarticletitle{Tweety holmes: A browser extension for abusive twitter profile detection}. In \bibinfo{booktitle}{\emph{Companion of the 2018 ACM Conference on Computer Supported Cooperative Work and Social Computing}}. \bibinfo{pages}{17--20}.
\newblock


\bibitem[\protect\citeauthoryear{Lenhart, Ybarra, Zickuhr, and Price-Feeney}{Lenhart et~al\mbox{.}}{2016}]%
        {lenhart2016online}
\bibfield{author}{\bibinfo{person}{Amanda Lenhart}, \bibinfo{person}{Michele Ybarra}, \bibinfo{person}{Kathryn Zickuhr}, {and} \bibinfo{person}{Myeshia Price-Feeney}.} \bibinfo{year}{2016}\natexlab{}.
\newblock \bibinfo{booktitle}{\emph{Online harassment, digital abuse, and cyberstalking in America}}.
\newblock \bibinfo{publisher}{Data and Society Research Institute}.
\newblock


\bibitem[\protect\citeauthoryear{Lewis, Zamith, and Coddington}{Lewis et~al\mbox{.}}{2020}]%
        {lewis2020online}
\bibfield{author}{\bibinfo{person}{Seth~C Lewis}, \bibinfo{person}{Rodrigo Zamith}, {and} \bibinfo{person}{Mark Coddington}.} \bibinfo{year}{2020}\natexlab{}.
\newblock \showarticletitle{Online harassment and its implications for the journalist--audience relationship}.
\newblock \bibinfo{journal}{\emph{Digital Journalism}} \bibinfo{volume}{8}, \bibinfo{number}{8} (\bibinfo{year}{2020}), \bibinfo{pages}{1047--1067}.
\newblock


\bibitem[\protect\citeauthoryear{Lowd}{Lowd}{2018a}]%
        {lowd2018can}
\bibfield{author}{\bibinfo{person}{Daniel Lowd}.} \bibinfo{year}{2018}\natexlab{a}.
\newblock \showarticletitle{Can Facebook use AI to fight online abuse}.
\newblock \bibinfo{journal}{\emph{Scientific American}} (\bibinfo{year}{2018}).
\newblock


\bibitem[\protect\citeauthoryear{Lowd}{Lowd}{2018b}]%
        {lowd2018}
\bibfield{author}{\bibinfo{person}{Daniel Lowd}.} \bibinfo{year}{2018}\natexlab{b}.
\newblock \bibinfo{title}{Can Facebook Use AI to Fight Online Abuse?}
\newblock
\newblock
\urldef\tempurl%
\url{https://www.scientificamerican.com/article/can-facebook-use-ai-to-fight-online-abuse/}
\showURL{%
Retrieved September 9, 2021 from \tempurl}


\bibitem[\protect\citeauthoryear{Lumsden and Morgan}{Lumsden and Morgan}{2017}]%
        {lumsden2017media}
\bibfield{author}{\bibinfo{person}{Karen Lumsden} {and} \bibinfo{person}{Heather Morgan}.} \bibinfo{year}{2017}\natexlab{}.
\newblock \showarticletitle{Media framing of trolling and online abuse: silencing strategies, symbolic violence, and victim blaming}.
\newblock \bibinfo{journal}{\emph{Feminist Media Studies}} \bibinfo{volume}{17}, \bibinfo{number}{6} (\bibinfo{year}{2017}), \bibinfo{pages}{926--940}.
\newblock


\bibitem[\protect\citeauthoryear{Mahar, Zhang, and Karger}{Mahar et~al\mbox{.}}{2018}]%
        {mahar2018squadbox}
\bibfield{author}{\bibinfo{person}{Kaitlin Mahar}, \bibinfo{person}{Amy~X Zhang}, {and} \bibinfo{person}{David Karger}.} \bibinfo{year}{2018}\natexlab{}.
\newblock \showarticletitle{Squadbox: A tool to combat email harassment using friendsourced moderation}. In \bibinfo{booktitle}{\emph{Proceedings of the 2018 CHI Conference on Human Factors in Computing Systems}}. \bibinfo{pages}{1--13}.
\newblock


\bibitem[\protect\citeauthoryear{Marshak}{Marshak}{2017}]%
        {marshak2017online}
\bibfield{author}{\bibinfo{person}{Emma Marshak}.} \bibinfo{year}{2017}\natexlab{}.
\newblock \showarticletitle{Online harassment: A legislative solution}.
\newblock \bibinfo{journal}{\emph{Harv. J. on Legis.}}  \bibinfo{volume}{54} (\bibinfo{year}{2017}), \bibinfo{pages}{503}.
\newblock


\bibitem[\protect\citeauthoryear{Mathew, Saha, Tharad, Rajgaria, Singhania, Maity, Goyal, and Mukherjee}{Mathew et~al\mbox{.}}{2019}]%
        {mathew2019thou}
\bibfield{author}{\bibinfo{person}{Binny Mathew}, \bibinfo{person}{Punyajoy Saha}, \bibinfo{person}{Hardik Tharad}, \bibinfo{person}{Subham Rajgaria}, \bibinfo{person}{Prajwal Singhania}, \bibinfo{person}{Suman~Kalyan Maity}, \bibinfo{person}{Pawan Goyal}, {and} \bibinfo{person}{Animesh Mukherjee}.} \bibinfo{year}{2019}\natexlab{}.
\newblock \showarticletitle{Thou shalt not hate: Countering online hate speech}. In \bibinfo{booktitle}{\emph{Proceedings of the international AAAI conference on web and social media}}, Vol.~\bibinfo{volume}{13}. \bibinfo{pages}{369--380}.
\newblock


\bibitem[\protect\citeauthoryear{Mishra, Yannakoudakis, and Shutova}{Mishra et~al\mbox{.}}{2019}]%
        {mishra2019tackling}
\bibfield{author}{\bibinfo{person}{Pushkar Mishra}, \bibinfo{person}{Helen Yannakoudakis}, {and} \bibinfo{person}{Ekaterina Shutova}.} \bibinfo{year}{2019}\natexlab{}.
\newblock \showarticletitle{Tackling online abuse: A survey of automated abuse detection methods}.
\newblock \bibinfo{journal}{\emph{arXiv preprint arXiv:1908.06024}} (\bibinfo{year}{2019}).
\newblock


\bibitem[\protect\citeauthoryear{Moitra, Ahmed, and Chandra}{Moitra et~al\mbox{.}}{2021}]%
        {moitra2021parsing}
\bibfield{author}{\bibinfo{person}{Aparna Moitra}, \bibinfo{person}{Syed~Ishtiaque Ahmed}, {and} \bibinfo{person}{Priyank Chandra}.} \bibinfo{year}{2021}\natexlab{}.
\newblock \showarticletitle{Parsing the'Me'in\# MeToo: Sexual Harassment, Social Media, and Justice Infrastructures}.
\newblock \bibinfo{journal}{\emph{Proceedings of the ACM on Human-Computer Interaction}} \bibinfo{volume}{5}, \bibinfo{number}{CSCW1} (\bibinfo{year}{2021}), \bibinfo{pages}{1--34}.
\newblock


\bibitem[\protect\citeauthoryear{Molina-Gil, Concepci{\'o}n-S{\'a}nchez, and Caballero-Gil}{Molina-Gil et~al\mbox{.}}{2019}]%
        {molina2019harassment}
\bibfield{author}{\bibinfo{person}{Jezabel Molina-Gil}, \bibinfo{person}{Jos{\'e}~A Concepci{\'o}n-S{\'a}nchez}, {and} \bibinfo{person}{Pino Caballero-Gil}.} \bibinfo{year}{2019}\natexlab{}.
\newblock \showarticletitle{Harassment detection using machine learning and fuzzy logic techniques}. In \bibinfo{booktitle}{\emph{Multidisciplinary digital publishing institute proceedings}}, Vol.~\bibinfo{volume}{31}. \bibinfo{pages}{27}.
\newblock


\bibitem[\protect\citeauthoryear{Nova, Rifat, Saha, Ahmed, and Guha}{Nova et~al\mbox{.}}{2019}]%
        {nova2019online}
\bibfield{author}{\bibinfo{person}{Fayika~Farhat Nova}, \bibinfo{person}{MD~Rashidujjaman Rifat}, \bibinfo{person}{Pratyasha Saha}, \bibinfo{person}{Syed~Ishtiaque Ahmed}, {and} \bibinfo{person}{Shion Guha}.} \bibinfo{year}{2019}\natexlab{}.
\newblock \showarticletitle{Online sexual harassment over anonymous social media in Bangladesh}. In \bibinfo{booktitle}{\emph{Proceedings of the Tenth International Conference on Information and Communication Technologies and Development}}. \bibinfo{pages}{1--12}.
\newblock


\bibitem[\protect\citeauthoryear{Ojanen, Boonmongkon, Samakkeekarom, Samoh, Cholratana, and Guadamuz}{Ojanen et~al\mbox{.}}{2015}]%
        {ojanen2015connections}
\bibfield{author}{\bibinfo{person}{Timo~Tapani Ojanen}, \bibinfo{person}{Pimpawun Boonmongkon}, \bibinfo{person}{Ronnapoom Samakkeekarom}, \bibinfo{person}{Nattharat Samoh}, \bibinfo{person}{Mudjalin Cholratana}, {and} \bibinfo{person}{Thomas~Ebanan Guadamuz}.} \bibinfo{year}{2015}\natexlab{}.
\newblock \showarticletitle{Connections between online harassment and offline violence among youth in Central Thailand}.
\newblock \bibinfo{journal}{\emph{Child abuse \& neglect}}  \bibinfo{volume}{44} (\bibinfo{year}{2015}), \bibinfo{pages}{159--169}.
\newblock


\bibitem[\protect\citeauthoryear{Posetti, Aboulez, Bontheva, Harrison, and Waisbord}{Posetti et~al\mbox{.}}{2020}]%
        {posetti2020online}
\bibfield{author}{\bibinfo{person}{Julie Posetti}, \bibinfo{person}{Nermine Aboulez}, \bibinfo{person}{K Bontheva}, \bibinfo{person}{Jackie Harrison}, {and} \bibinfo{person}{Silvio Waisbord}.} \bibinfo{year}{2020}\natexlab{}.
\newblock \bibinfo{title}{Online Violence Against Women Journalists: A Global Snapshot of Incidence and Impacts}.
\newblock
\newblock


\bibitem[\protect\citeauthoryear{Posetti, Harrison, and Waisbord}{Posetti et~al\mbox{.}}{2017}]%
        {posetti2017conversation}
\bibfield{author}{\bibinfo{person}{Julie Posetti}, \bibinfo{person}{Jackie Harrison}, {and} \bibinfo{person}{Silvio Waisbord}.} \bibinfo{year}{2017}\natexlab{}.
\newblock \bibinfo{title}{Online attacks on female journalists are increasingly spilling into the ‘real world’ – new research}.
\newblock
\newblock
\urldef\tempurl%
\url{https://theconversation.com/online-attacks-on-female-journalists-are-increasingly-spilling-into-the-real-world-new-research-150791}
\showURL{%
Retrieved September 9, 2021 from \tempurl}


\bibitem[\protect\citeauthoryear{Posetti and Storm}{Posetti and Storm}{2018}]%
        {posetti2018violence}
\bibfield{author}{\bibinfo{person}{Julie Posetti} {and} \bibinfo{person}{Hanna Storm}.} \bibinfo{year}{2018}\natexlab{}.
\newblock \showarticletitle{Violence Against Women Journalists—Online and Offline}.
\newblock \bibinfo{journal}{\emph{Setting the Gender Agenda for Communication Policy: New Proposals from the Global Alliance on Media and Gender}} (\bibinfo{year}{2018}), \bibinfo{pages}{75--86}.
\newblock


\bibitem[\protect\citeauthoryear{Redmiles, Bodford, and Blackwell}{Redmiles et~al\mbox{.}}{2019}]%
        {redmiles2019just}
\bibfield{author}{\bibinfo{person}{Elissa~M Redmiles}, \bibinfo{person}{Jessica Bodford}, {and} \bibinfo{person}{Lindsay Blackwell}.} \bibinfo{year}{2019}\natexlab{}.
\newblock \showarticletitle{“I just want to feel safe”: A Diary Study of Safety Perceptions on Social Media}. In \bibinfo{booktitle}{\emph{Proceedings of the International AAAI Conference on Web and Social Media}}, Vol.~\bibinfo{volume}{13}. \bibinfo{pages}{405--416}.
\newblock


\bibitem[\protect\citeauthoryear{Ringel and Davidson}{Ringel and Davidson}{2020}]%
        {ringel2020proactive}
\bibfield{author}{\bibinfo{person}{Sharon Ringel} {and} \bibinfo{person}{Roei Davidson}.} \bibinfo{year}{2020}\natexlab{}.
\newblock \showarticletitle{Proactive ephemerality: How journalists use automated and manual tweet deletion to minimize risk and its consequences for social media as a public archive}.
\newblock \bibinfo{journal}{\emph{new media \& society}} (\bibinfo{year}{2020}), \bibinfo{pages}{1461444820972389}.
\newblock


\bibitem[\protect\citeauthoryear{Sambasivan, Batool, Ahmed, Matthews, Thomas, Gayt{\'a}n-Lugo, Nemer, Bursztein, Churchill, and Consolvo}{Sambasivan et~al\mbox{.}}{2019}]%
        {sambasivan2019they}
\bibfield{author}{\bibinfo{person}{Nithya Sambasivan}, \bibinfo{person}{Amna Batool}, \bibinfo{person}{Nova Ahmed}, \bibinfo{person}{Tara Matthews}, \bibinfo{person}{Kurt Thomas}, \bibinfo{person}{Laura~Sanely Gayt{\'a}n-Lugo}, \bibinfo{person}{David Nemer}, \bibinfo{person}{Elie Bursztein}, \bibinfo{person}{Elizabeth Churchill}, {and} \bibinfo{person}{Sunny Consolvo}.} \bibinfo{year}{2019}\natexlab{}.
\newblock \showarticletitle{" They Don't Leave Us Alone Anywhere We Go" Gender and Digital Abuse in South Asia}. In \bibinfo{booktitle}{\emph{proceedings of the 2019 CHI Conference on Human Factors in Computing Systems}}. \bibinfo{pages}{1--14}.
\newblock


\bibitem[\protect\citeauthoryear{Scheuerman, Branham, and Hamidi}{Scheuerman et~al\mbox{.}}{2018}]%
        {scheuerman2018safe}
\bibfield{author}{\bibinfo{person}{Morgan~Klaus Scheuerman}, \bibinfo{person}{Stacy~M Branham}, {and} \bibinfo{person}{Foad Hamidi}.} \bibinfo{year}{2018}\natexlab{}.
\newblock \showarticletitle{Safe spaces and safe places: Unpacking technology-mediated experiences of safety and harm with transgender people}.
\newblock \bibinfo{journal}{\emph{Proceedings of the ACM on Human-Computer Interaction}} \bibinfo{volume}{2}, \bibinfo{number}{CSCW} (\bibinfo{year}{2018}), \bibinfo{pages}{1--27}.
\newblock


\bibitem[\protect\citeauthoryear{Schoenebeck, Haimson, and Nakamura}{Schoenebeck et~al\mbox{.}}{2021}]%
        {sarita2021}
\bibfield{author}{\bibinfo{person}{Sarita Schoenebeck}, \bibinfo{person}{Oliver~L Haimson}, {and} \bibinfo{person}{Lisa Nakamura}.} \bibinfo{year}{2021}\natexlab{}.
\newblock \showarticletitle{Drawing from justice theories to support targets of online harassment}.
\newblock \bibinfo{journal}{\emph{New Media \& Society}} \bibinfo{volume}{23}, \bibinfo{number}{5} (\bibinfo{year}{2021}), \bibinfo{pages}{1278--1300}.
\newblock
\urldef\tempurl%
\url{https://doi.org/10.1177/1461444820913122}
\showDOI{\tempurl}
\showeprint{https://doi.org/10.1177/1461444820913122}


\bibitem[\protect\citeauthoryear{Seering}{Seering}{2020}]%
        {seering2020reconsidering}
\bibfield{author}{\bibinfo{person}{Joseph Seering}.} \bibinfo{year}{2020}\natexlab{}.
\newblock \showarticletitle{Reconsidering Self-Moderation: the Role of Research in Supporting Community-Based Models for Online Content Moderation}.
\newblock \bibinfo{journal}{\emph{Proceedings of the ACM on Human-Computer Interaction}} \bibinfo{volume}{4}, \bibinfo{number}{CSCW2} (\bibinfo{year}{2020}), \bibinfo{pages}{1--28}.
\newblock


\bibitem[\protect\citeauthoryear{Seralathan}{Seralathan}{2016}]%
        {seralathan2016making}
\bibfield{author}{\bibinfo{person}{A~Meena Seralathan}.} \bibinfo{year}{2016}\natexlab{}.
\newblock \showarticletitle{Making the time fit the crime: Clearly defining online harassment crimes and providing incentives for investigating online threats in the digital age}.
\newblock \bibinfo{journal}{\emph{Brook. J. Int'l L.}}  \bibinfo{volume}{42} (\bibinfo{year}{2016}), \bibinfo{pages}{425}.
\newblock


\bibitem[\protect\citeauthoryear{Sim{\~o}es, Alcantara, and Carona}{Sim{\~o}es et~al\mbox{.}}{[n.\,d.]}]%
        {simoesonline}
\bibfield{author}{\bibinfo{person}{Rita~Bas{\'\i}lio Sim{\~o}es}, \bibinfo{person}{Juliana Alcantara}, {and} \bibinfo{person}{Liliana Carona}.} \bibinfo{year}{[n.\,d.]}\natexlab{}.
\newblock \showarticletitle{Online abuse against female journalists: A scoping review}.
\newblock  (\bibinfo{year}{[n.\,d.]}).
\newblock


\bibitem[\protect\citeauthoryear{Steinmetz}{Steinmetz}{2018}]%
        {steinmetz2020}
\bibfield{author}{\bibinfo{person}{Katy Steinmetz}.} \bibinfo{year}{2018}\natexlab{}.
\newblock \bibinfo{title}{How The Internet Can Make Hate Seem Normal — And Why That's So Dangerous}.
\newblock
\newblock
\urldef\tempurl%
\url{https://time.com/5439713/online-hate-speech/}
\showURL{%
Retrieved September 9, 2021 from \tempurl}


\bibitem[\protect\citeauthoryear{Steinmetz}{Steinmetz}{nd}]%
        {BlockParty}
\bibfield{author}{\bibinfo{person}{Katy Steinmetz}.} \bibinfo{year}{n.d.}\natexlab{}.
\newblock \bibinfo{title}{BlockParty}.
\newblock
\newblock
\urldef\tempurl%
\url{https://www.blockpartyapp.com/}
\showURL{%
Retrieved September 9, 2021 from \tempurl}


\bibitem[\protect\citeauthoryear{Stevens, Nurse, and Arief}{Stevens et~al\mbox{.}}{2021}]%
        {stevens2021cyber}
\bibfield{author}{\bibinfo{person}{Francesca Stevens}, \bibinfo{person}{Jason~RC Nurse}, {and} \bibinfo{person}{Budi Arief}.} \bibinfo{year}{2021}\natexlab{}.
\newblock \showarticletitle{Cyber stalking, cyber harassment, and adult mental health: A systematic review}.
\newblock \bibinfo{journal}{\emph{Cyberpsychology, Behavior, and Social Networking}} \bibinfo{volume}{24}, \bibinfo{number}{6} (\bibinfo{year}{2021}), \bibinfo{pages}{367--376}.
\newblock


\bibitem[\protect\citeauthoryear{Sugiura and Smith}{Sugiura and Smith}{2020}]%
        {sugiura2020victim}
\bibfield{author}{\bibinfo{person}{Lisa Sugiura} {and} \bibinfo{person}{April Smith}.} \bibinfo{year}{2020}\natexlab{}.
\newblock \showarticletitle{Victim blaming, responsibilization and resilience in online sexual abuse and harassment}.
\newblock In \bibinfo{booktitle}{\emph{Victimology}}. \bibinfo{publisher}{Springer}, \bibinfo{pages}{45--79}.
\newblock


\bibitem[\protect\citeauthoryear{Sultana, Deb, Bhattacharjee, Hasan, Alam, Chakraborty, Roy, Ahmed, Moitra, Amin, et~al\mbox{.}}{Sultana et~al\mbox{.}}{2021}]%
        {sultana2021unmochon}
\bibfield{author}{\bibinfo{person}{Sharifa Sultana}, \bibinfo{person}{Mitrasree Deb}, \bibinfo{person}{Ananya Bhattacharjee}, \bibinfo{person}{Shaid Hasan}, \bibinfo{person}{SM~Raihanul Alam}, \bibinfo{person}{Trishna Chakraborty}, \bibinfo{person}{Prianka Roy}, \bibinfo{person}{Samira~Fairuz Ahmed}, \bibinfo{person}{Aparna Moitra}, \bibinfo{person}{M~Ashraful Amin}, {et~al\mbox{.}}} \bibinfo{year}{2021}\natexlab{}.
\newblock \showarticletitle{‘Unmochon’: A Tool to Combat Online Sexual Harassment over Facebook Messenger}. In \bibinfo{booktitle}{\emph{Proceedings of the 2021 CHI Conference on Human Factors in Computing Systems}}. \bibinfo{pages}{1--18}.
\newblock


\bibitem[\protect\citeauthoryear{Thomas, Akhawe, Bailey, Boneh, Bursztein, Consolvo, Dell, Durumeric, Kelley, Kumar, et~al\mbox{.}}{Thomas et~al\mbox{.}}{2021}]%
        {thomas2021sok}
\bibfield{author}{\bibinfo{person}{Kurt Thomas}, \bibinfo{person}{Devdatta Akhawe}, \bibinfo{person}{Michael Bailey}, \bibinfo{person}{Dan Boneh}, \bibinfo{person}{Elie Bursztein}, \bibinfo{person}{Sunny Consolvo}, \bibinfo{person}{Nicola Dell}, \bibinfo{person}{Zakir Durumeric}, \bibinfo{person}{Patrick~Gage Kelley}, \bibinfo{person}{Deepak Kumar}, {et~al\mbox{.}}} \bibinfo{year}{2021}\natexlab{}.
\newblock \showarticletitle{Sok: Hate, harassment, and the changing landscape of online abuse}.
\newblock  (\bibinfo{year}{2021}).
\newblock


\bibitem[\protect\citeauthoryear{UNESCO}{UNESCO}{2021}]%
        {UNESCO2021}
\bibfield{author}{\bibinfo{person}{UNESCO}.} \bibinfo{year}{2021}\natexlab{}.
\newblock \bibinfo{title}{Freedom of expression and the safety of foreign correspondents: trends, challenges and responses}.
\newblock
\newblock
\urldef\tempurl%
\url{https://unesdoc.unesco.org/ark:/48223/pf0000378300}
\showURL{%
Retrieved September 9, 2021 from \tempurl}


\bibitem[\protect\citeauthoryear{Van~Hee, Jacobs, Emmery, Desmet, Lefever, Verhoeven, De~Pauw, Daelemans, and Hoste}{Van~Hee et~al\mbox{.}}{2018}]%
        {van2018automatic}
\bibfield{author}{\bibinfo{person}{Cynthia Van~Hee}, \bibinfo{person}{Gilles Jacobs}, \bibinfo{person}{Chris Emmery}, \bibinfo{person}{Bart Desmet}, \bibinfo{person}{Els Lefever}, \bibinfo{person}{Ben Verhoeven}, \bibinfo{person}{Guy De~Pauw}, \bibinfo{person}{Walter Daelemans}, {and} \bibinfo{person}{V{\'e}ronique Hoste}.} \bibinfo{year}{2018}\natexlab{}.
\newblock \showarticletitle{Automatic detection of cyberbullying in social media text}.
\newblock \bibinfo{journal}{\emph{PloS one}} \bibinfo{volume}{13}, \bibinfo{number}{10} (\bibinfo{year}{2018}), \bibinfo{pages}{e0203794}.
\newblock


\bibitem[\protect\citeauthoryear{Vashistha, Garg, Anderson, and Raza}{Vashistha et~al\mbox{.}}{2019}]%
        {vashistha2019threats}
\bibfield{author}{\bibinfo{person}{Aditya Vashistha}, \bibinfo{person}{Abhinav Garg}, \bibinfo{person}{Richard Anderson}, {and} \bibinfo{person}{Agha~Ali Raza}.} \bibinfo{year}{2019}\natexlab{}.
\newblock \showarticletitle{Threats, abuses, flirting, and blackmail: Gender inequity in social media voice forums}. In \bibinfo{booktitle}{\emph{Proceedings of the 2019 CHI Conference on Human Factors in Computing Systems}}. \bibinfo{pages}{1--13}.
\newblock


\bibitem[\protect\citeauthoryear{Veletsianos, Houlden, Hodson, and Gosse}{Veletsianos et~al\mbox{.}}{2018}]%
        {veletsianos2018women}
\bibfield{author}{\bibinfo{person}{George Veletsianos}, \bibinfo{person}{Shandell Houlden}, \bibinfo{person}{Jaigris Hodson}, {and} \bibinfo{person}{Chandell Gosse}.} \bibinfo{year}{2018}\natexlab{}.
\newblock \showarticletitle{Women scholars’ experiences with online harassment and abuse: Self-protection, resistance, acceptance, and self-blame}.
\newblock \bibinfo{journal}{\emph{New Media \& Society}} \bibinfo{volume}{20}, \bibinfo{number}{12} (\bibinfo{year}{2018}), \bibinfo{pages}{4689--4708}.
\newblock


\bibitem[\protect\citeauthoryear{Vincent}{Vincent}{2020}]%
        {vincent2020}
\bibfield{author}{\bibinfo{person}{James Vincent}.} \bibinfo{year}{2020}\natexlab{}.
\newblock \bibinfo{title}{More Americans are being harassed online because of their race, religion, or sexuality}.
\newblock
\newblock
\urldef\tempurl%
\url{https://www.theverge.com/2020/6/23/21300127/online-harassment-2020-adl-survey-race-religion-sexuality}
\showURL{%
Retrieved September 9, 2021 from \tempurl}


\bibitem[\protect\citeauthoryear{Vitak, Chadha, Steiner, and Ashktorab}{Vitak et~al\mbox{.}}{2017}]%
        {vitak2017identifying}
\bibfield{author}{\bibinfo{person}{Jessica Vitak}, \bibinfo{person}{Kalyani Chadha}, \bibinfo{person}{Linda Steiner}, {and} \bibinfo{person}{Zahra Ashktorab}.} \bibinfo{year}{2017}\natexlab{}.
\newblock \showarticletitle{Identifying women's experiences with and strategies for mitigating negative effects of online harassment}. In \bibinfo{booktitle}{\emph{Proceedings of the 2017 ACM Conference on Computer Supported Cooperative Work and Social Computing}}. \bibinfo{pages}{1231--1245}.
\newblock


\bibitem[\protect\citeauthoryear{Wijesiriwardene, Inan, Kursuncu, Gaur, Shalin, Thirunarayan, Sheth, and Arpinar}{Wijesiriwardene et~al\mbox{.}}{2020}]%
        {wijesiriwardene2020alone}
\bibfield{author}{\bibinfo{person}{Thilini Wijesiriwardene}, \bibinfo{person}{Hale Inan}, \bibinfo{person}{Ugur Kursuncu}, \bibinfo{person}{Manas Gaur}, \bibinfo{person}{Valerie~L Shalin}, \bibinfo{person}{Krishnaprasad Thirunarayan}, \bibinfo{person}{Amit Sheth}, {and} \bibinfo{person}{I~Budak Arpinar}.} \bibinfo{year}{2020}\natexlab{}.
\newblock \showarticletitle{Alone: A dataset for toxic behavior among adolescents on twitter}. In \bibinfo{booktitle}{\emph{International Conference on Social Informatics}}. Springer, \bibinfo{pages}{427--439}.
\newblock


\bibitem[\protect\citeauthoryear{Williams, Burnap, Javed, Liu, and Ozalp}{Williams et~al\mbox{.}}{2020}]%
        {williams2020hate}
\bibfield{author}{\bibinfo{person}{Matthew~L Williams}, \bibinfo{person}{Pete Burnap}, \bibinfo{person}{Amir Javed}, \bibinfo{person}{Han Liu}, {and} \bibinfo{person}{Sefa Ozalp}.} \bibinfo{year}{2020}\natexlab{}.
\newblock \showarticletitle{Hate in the machine: Anti-Black and anti-Muslim social media posts as predictors of offline racially and religiously aggravated crime}.
\newblock \bibinfo{journal}{\emph{The British Journal of Criminology}} \bibinfo{volume}{60}, \bibinfo{number}{1} (\bibinfo{year}{2020}), \bibinfo{pages}{93--117}.
\newblock


\bibitem[\protect\citeauthoryear{Women}{Women}{2020}]%
        {women2020online}
\bibfield{author}{\bibinfo{person}{UN Women}.} \bibinfo{year}{2020}\natexlab{}.
\newblock \showarticletitle{Online and ICT facilitated violence against women and girls during COVID-19}. New York: UN Women. https://www. itu. int/net4/wsis/forum/2020/Files/talkx~….
\newblock


\bibitem[\protect\citeauthoryear{Woodruff}{Woodruff}{2014}]%
        {woodruff2014necessary}
\bibfield{author}{\bibinfo{person}{Allison Woodruff}.} \bibinfo{year}{2014}\natexlab{}.
\newblock \showarticletitle{Necessary, unpleasant, and disempowering: Reputation management in the internet age}. In \bibinfo{booktitle}{\emph{Proceedings of the SIGCHI conference on human factors in computing systems}}. \bibinfo{pages}{149--158}.
\newblock


\bibitem[\protect\citeauthoryear{Yoo and Dourish}{Yoo and Dourish}{2021}]%
        {yoo2021anshimi}
\bibfield{author}{\bibinfo{person}{Chaeyoon Yoo} {and} \bibinfo{person}{Paul Dourish}.} \bibinfo{year}{2021}\natexlab{}.
\newblock \showarticletitle{Anshimi: Women's Perceptions of Safety Data and the Efficacy of a Safety Application in Seoul}.
\newblock \bibinfo{journal}{\emph{Proceedings of the ACM on Human-Computer Interaction}} \bibinfo{volume}{5}, \bibinfo{number}{CSCW1} (\bibinfo{year}{2021}), \bibinfo{pages}{1--21}.
\newblock


\bibitem[\protect\citeauthoryear{Younas, Naseem, and Mustafa}{Younas et~al\mbox{.}}{2020}]%
        {younas2020patriarchy}
\bibfield{author}{\bibinfo{person}{Fouzia Younas}, \bibinfo{person}{Mustafa Naseem}, {and} \bibinfo{person}{Maryam Mustafa}.} \bibinfo{year}{2020}\natexlab{}.
\newblock \showarticletitle{Patriarchy and social media: Women only facebook groups as safe spaces for support seeking in Pakistan}. In \bibinfo{booktitle}{\emph{Proceedings of the 2020 International Conference on Information and Communication Technologies and Development}}. \bibinfo{pages}{1--11}.
\newblock


\bibitem[\protect\citeauthoryear{Zheng, Zhang, and Thing}{Zheng et~al\mbox{.}}{2019}]%
        {zheng2019survey}
\bibfield{author}{\bibinfo{person}{Lilei Zheng}, \bibinfo{person}{Ying Zhang}, {and} \bibinfo{person}{Vrizlynn~LL Thing}.} \bibinfo{year}{2019}\natexlab{}.
\newblock \showarticletitle{A survey on image tampering and its detection in real-world photos}.
\newblock \bibinfo{journal}{\emph{Journal of Visual Communication and Image Representation}}  \bibinfo{volume}{58} (\bibinfo{year}{2019}), \bibinfo{pages}{380--399}.
\newblock


\bibitem[\protect\citeauthoryear{Zone}{Zone}{2016}]%
        {jigsawmedium}
\bibfield{author}{\bibinfo{person}{Jigsaw Medium~Blog Zone}.} \bibinfo{year}{2016}\natexlab{}.
\newblock  (\bibinfo{year}{2016}).
\newblock
\urldef\tempurl%
\url{https://medium.com/jigsaw/designing-for-an-at-risk-world-75081c6fa061}
\showURL{%
\tempurl}


\end{thebibliography}

\clearpage
\appendix
\includepdf{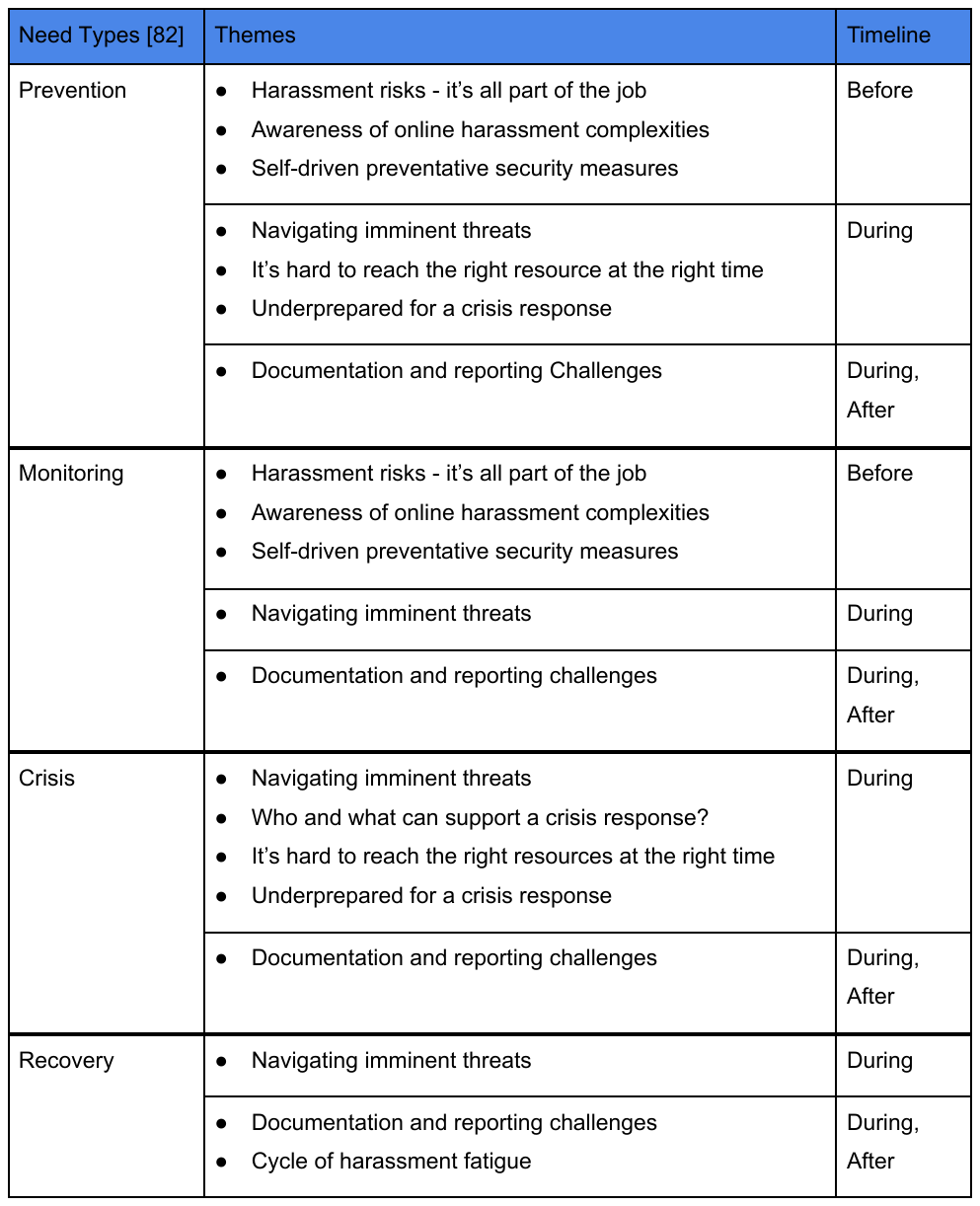}


\end{document}